\documentclass[aps,pre,preprint,groupedaddress,showpacs]{revtex4-1}
\usepackage[dvips]{graphicx}
\usepackage{textcomp}
\usepackage{amssymb}
\usepackage{dsdshorthand}

\begin{document}

\title{
\Large\bf Two- and three-point functions at criticality: \\
 Monte Carlo simulations of the improved three-dimensional Blume-Capel model
}

\author{Martin Hasenbusch}
\email[]{Martin.Hasenbusch@physik.hu-berlin.de}
\affiliation{
Institut f\"ur Physik, Humboldt-Universit\"at zu Berlin,
Newtonstr. 15, 12489 Berlin, Germany}

\date{\today}

\begin{abstract}
We compute two- and three-point functions at criticality for the 
three-dimensional
Ising universality class. To this end we simulate the improved Blume-Capel 
model at the critical temperature on lattices of a linear size up to 
$L=1600$. As check also
simulations of the spin-1/2 Ising model are performed. We find 
$f_{\sigma \sigma \epsilon} = 1.051(1)$ and 
$f_{\epsilon \epsilon \epsilon} =1.533(5)$
for operator product expansion coefficients. 
These results are consistent with but less precise
than those recently obtained by using the bootstrap method. 
An important ingredient in our simulations
is a variance reduced estimator of $N$-point functions. Finite size 
corrections vanish with $L^{-\Delta_{\epsilon}}$, where $L$ is the linear 
size of the lattice and $\Delta_{\epsilon}$ is the scaling dimension of
the leading $Z_2$-even scalar $\epsilon$. 
\end{abstract}

\pacs{05.50.+q, 05.70.Jk, 05.10.Ln}
\keywords{}
\maketitle

\section{Introduction}
Recently we have seen enormous progress in the understanding of
critical phenomena in three dimensions by using the conformal
bootstrap approach \cite{Kos:2014bka,Gliozzi:2014jsa,Kos:2016ysd,Simmons-Duffin:2016wlq}, 
to give only a few references. For a recent lecture note on the subject see
\cite{Simmons-Duffin:2016gjk}. 
In particular,  precise
numbers for scaling dimensions and operator product expansion (OPE) 
coefficients  for the
universality class of the three-dimensional Ising model were
obtained. For a very detailed account see ref. \cite{Simmons-Duffin:2016wlq}.

The functional form of two-point 
functions of primary operators is fixed by conformal invariance
\begin{equation}
\label{twopoint}
 \langle  \cO_1(x_1)  \cO_2(x_2)  \rangle =
  \frac{C_1 \delta_{\Delta_1,\Delta_2}}{|x_1 -  x_2|^{2 \Delta_1}} \;\;,
\end{equation}
where $\cO_i$ is the operator taken at the site $x_i$ and
$\Delta_i$ is its scaling dimension. 
The scaling dimensions of the leading $Z_2$-odd scalar $\sigma$ and 
the leading $Z_2$-even scalar $\epsilon$ for the three-dimensional Ising 
universality class are 
$\Delta_{\sigma} = 0.5181489(10)$ and $\Delta_{\epsilon} = 1.412625(10)$,
respectively \cite{Kos:2016ysd,Simmons-Duffin:2016wlq}.  
The critical exponents that are usually discussed in critical phenomena
can be deduced from these two scaling dimensions. Let us consider two 
examples:
The exponent $\eta= d + 2 -2 y_h = d + 2 -2 (d -\Delta_{\sigma})=
2 \Delta_{\sigma}-d +2 = 0.0362978(20)$
governs the decay of the spin-spin correlation function at the critical point,
 where 
$y_h$ is the RG-exponent related to the external field and $d$ is the 
dimension of the system.
The critical exponent of the correlation length is given by
$\nu=1/y_t=1/(d-\Delta_{\epsilon})=0.6299709(40)$, where $y_t$ is the
RG-exponent related to the temperature. In table 2 of 
\cite{Simmons-Duffin:2016wlq} one finds
$\omega= \Delta_{\epsilon'}-3 = 0.82968(23)$  for the exponent of the
leading correction. These results are by far  more accurate than 
previous ones obtained by other methods. For example using 
Monte Carlo simulations of the improved Blume-Capel model 
$\nu=0.63002(10)$, $\eta=0.03627(10)$, and $\omega=0.832(6)$ had been 
obtained \cite{Hasenbusch:2011yya}. Results obtained from high-temperature 
series expansions, see for example 
refs. \cite{Campostrini:2002cf,Butera:2005zf}, are slightly less accurate 
than those from Monte Carlo simulations.  Field theoretic methods, 
see for example ref. \cite{Guida:1998bx}, give even less precise 
estimates. In our numerical study, corrections caused by the breaking 
of the rotational symmetry are an important issue. These are governed by 
the correction exponent $\omega_{NR}$. In table I of ref. 
\cite{Campostrini:2002cf} the authors quote $\omega_{NR}=2.0208(12)$, 
which is in reasonable agreement with $\omega_{NR}=2.022665(28)$ that 
follows from $\Delta=5.022665(28)$ for angular momentum $l=4$ given in 
table 2 of ref. \cite{Simmons-Duffin:2016wlq}. 
There is a very rich literature on critical phenomena, which 
we can not recapitulate here. We refer the reader to reviews of the 
subject \cite{Wilson:1973jj,Fisher:1974uq,Fisher:1998kv,Pelissetto:2000ek}.

Now let us turn to the OPE coefficients.
Also the form of three-point 
functions is fixed by conformal invariance.
Normalizing the operators such that $C_i=1$, eq.~(\ref{twopoint}), 
one gets \cite{Polyakov:1970xd}
\begin{equation}
\label{threepoint}
\langle  \cO_1(x_1)  \cO_2(x_2)   \cO_3(x_3) \rangle
= \frac{f_{123}}{
       |x_1-x_2|^{\Delta_1+\Delta_2-\Delta_3}
       |x_2-x_3|^{\Delta_2+\Delta_3-\Delta_1}
       |x_3-x_1|^{\Delta_3+\Delta_1-\Delta_2}
                } \;\;,
\end{equation}
where the OPE coefficients $f_{123}$   depend on the universality class.
Using 
the conformal bootstrap method, highly accurate results for OPE coefficients
of the Ising universality class in three dimensions were obtained
\cite{Kos:2016ysd,Simmons-Duffin:2016wlq}:
\begin{equation}
\label{fbssse}
f_{\sigma \sigma \epsilon} = 1.0518537(41)
\end{equation}
and
\begin{equation}
\label{fbseee}
f_{\epsilon \epsilon \epsilon} = 1.532435(19) \;\;.
\end{equation}
Until recently, there had been no results obtained by other methods
that could be compared with eqs.~(\ref{fbssse},\ref{fbseee}). In particular
in Monte Carlo simulations of lattice models, it is unclear how an infinite
volume at the critical point could be well approximated.

In ref. \cite{Caselle:2015csa} the requirement of an infinite volume was
circumvented by studying two-point correlators at the critical temperature 
applying a finite external field $h$. The OPE coefficients are obtained from 
the $h$-dependence of the two-point correlators. 
For details of the method we refer the reader 
to ref. \cite{Caselle:2015csa}.  Simulating the Ising model on a simple cubic
lattice, the authors find
\begin{equation}
f_{\sigma \sigma \epsilon} = 1.07(3)  \;\;,\; 
f_{\epsilon \epsilon \epsilon} = 1.45(30) \;\;. 
\end{equation}
These estimates were improved by using a trapping potential in ref. 
\cite{Costagliola:2015ier}:
\begin{equation}
f_{\sigma \sigma \epsilon} = 1.051(3)  \;\;,\; 
f_{\epsilon \epsilon \epsilon} = 1.32(15)  \;\;.
\end{equation}

Very recently, Herdeiro \cite{Herdeiro:2017jmv} computed two- and three-point
functions for the Ising model on the simple cubic lattice at the 
critical point, using the so called UV-sampler method.
The method had been tested at
the example of the Ising model on the square lattice \cite{Herdeiro:2016agy}. 
Fitting his data for the three-point 
function with a one parameter Ansatz,
fixing $\Delta_{\sigma}$ and $\Delta_{\epsilon}$ to their bootstrap values,
Herdeiro gets
\begin{equation}
f_{\sigma \sigma \epsilon} = 1.05334(203) \;\;,\; 
f_{\epsilon \epsilon \epsilon}  = 1.578(27) \;\;.
\end{equation}

Here we shall follow a different strategy. We simulated the improved
Blume-Capel model at the critical point on a simple cubic lattice with 
periodic boundary conditions. To keep finite size effects small,
we simulated  lattices of a linear size up to $L=1600$. Furthermore 
an extrapolation in the lattice size is performed.
It turns out that finite size effects at the critical point are 
$\propto L^{-\Delta_{\epsilon}}$ for all quantities that we study here.
We employ variance reduction as proposed in refs. 
\cite{Parisi:1983hm,Luscher:2001up}. This is in particular helpful in the case 
of the $\epsilon \epsilon \epsilon$ function.

The outline of the paper is the following: In the next section we recall 
the definition of the Blume-Capel model and summarize briefly the results
of ref. \cite{Hasenbusch:2011yya}. Then we define the two- and three-point
functions that we measure. Then in section \ref{fse}  
we discuss the finite size scaling 
behaviour of the quantities that we study. 
In sections \ref{variance} we discuss the application of the  
variance reduction method to our problem. Next we discuss the simulations 
that we perform. It follows the analysis of the data. Finally we conclude and 
give an outlook.

\section{The model}
As in previous work, 
we study the Blume-Capel model on the simple cubic lattice.
For a vanishing external field, it is defined by the reduced Hamiltonian
\begin{equation}
\label{BlumeCapel}
H = -\beta \sum_{<xy>}  s_x s_y
  + D \sum_x s_x^2   \;\; ,
\end{equation}
where the spin might assume the values $s_x \in \{-1, 0, 1 \}$. 
$x=(x^{(0)},x^{(1)},x^{(2)})$
denotes a site on the simple cubic lattice, where 
$x^{(i)} \in \{0,1,...,L_i-1\}$
and $<xy>$ denotes a pair of nearest neighbours on the lattice.
In this study we consider $L_0=L_1=L_2=L$ throughout.
The inverse temperature is denoted by $\beta=1/k_B T$. The partition function
is given by $Z = \sum_{\{s\}} \exp(- H)$, where the sum runs over all spin
configurations. The parameter $D$ controls the
density of vacancies $s_x=0$. In the limit $D \rightarrow - \infty$
vacancies are completely suppressed and hence the spin-1/2 Ising
model is recovered.

In  $d\ge 2$  dimensions the model undergoes a continuous phase transition
for $-\infty \le  D   < D_{tri} $ at a $\beta_c$ that depends on $D$, while
for $D > D_{tri}$ the model undergoes a first order phase transition,
where $D_{tri}=2.0313(4)$ for $d=3$, see ref. \cite{DeBl04}.

Numerically, using Monte Carlo simulations it has been shown that there
is a point $(D^*,\beta_c(D^*))$
on the line of second order phase transitions, where the amplitude
of leading corrections to scaling vanishes.
We refer to the Blume-Capel model
at values of $D$ that are good numerical approximations of $D^*$ as
improved Blume-Capel model. 
 For a more general discussion of improved models
see for example section 3.5 of \cite{myhabil} or section 2.3.1 of \cite{Pelissetto:2000ek}.
In \cite{Hasenbusch:2011yya} we
simulated the model at $D=0.655$ close to $\beta_c$ on lattices of a
linear size up to $L=360$. We obtained $\beta_c(0.655)=0.387721735(25)$
and $D^*=0.656(20)$.
The amplitude of leading corrections to scaling at $D=0.655$ is at
least by a factor of $30$ smaller than for the spin-1/2 Ising model.

Here we simulate the Blume-Capel model at $D=0.655$. 
Most of our simulations are performed at  
$\beta=0.387721735$. In order to check the sensitivity of the results
on $\beta$, we performed in addition a few simulations 
at $\beta=0.38772$ and $0.38772347$.  
In order to check the effect of leading corrections to scaling we 
also simulated the spin-1/2 Ising model on the simple cubic lattice
at $\beta=0.22165462$. Note that in eq.~(A2) of 
ref. \cite{Hasenbusch:2012spc} we quote $\beta_c=0.22165462(2)$.

\section{The observables}
On the lattice we identify
\begin{eqnarray}
\label{epsi}
 \epsilon(x)  &=& s_x^2 - \langle s_x^2 \rangle + ... \\
\label{sigi}
 \sigma(x)   &=& s_x + ... \;,
\end{eqnarray}
where in our numerical study $\langle s_x^2 \rangle$ is replaced by
its estimate obtained from the given simulation at finite $L$.
Corrections are caused by fields with the same symmetry properties 
but higher dimensions.

In the case of the Ising model, eq.~(\ref{epsi}) makes no sense. In the 
literature $s_x s_{y}$, where $x$ and $y$ are nearest neighbours,
is used instead of $s_x^2$.  Here, motivated by eq.~(\ref{sum2}) below, 
we used $S_x^2$, where $S_x = \sum_{y.nn.x} s_y$, 
where $y.nn.x$ means that $y$ is a nearest neighbour of $x$.

In order to keep the study tractable, we have to single out
a few directions for the displacements between the points. 
In the case of the two-point functions we consider displacements
along the axes
\begin{equation}
x_2-x_1 = (j,0,0) \; \;,\; x_2-x_1 = (0,j,0)  \; \;,\; x_2-x_1 = (0,0,j) \;,
\end{equation}
the face diagonals
\begin{eqnarray}
x_2-x_1 = (j,j,0) \; \;,\; x_2-x_1 &=& (j,0,j)  \; \;,\; x_2-x_1 = (0,j,j) \;,
\nonumber \\
x_2-x_1 = (j,-j,0) \; \;,\; x_2-x_1 &=& (j,0,-j) \; \;,\; x_2-x_1 = (0,j,-j) \;,
\end{eqnarray}
and space diagonals
\begin{equation}
x_2-x_1 = (j,j,j)  \; \;,\; x_2-x_1 = (j,j,-j)  \; \;,\; x_2-x_1 = (j,-j,j) \;,
\;\;x_2-x_1 = (j,-j,-j) \;,
\end{equation}
where $j$ is an integer. 
In the following we shall indicate these three directions by axis ($a$), 
face diagonal ($f$), and space diagonal ($d$), respectively.
In our simulation program we summed over all choices that are related by
symmetry to reduce the statistical error. In particular, we summed over all 
possible choices of $x_1$. In the following we shall denote the two-point 
function by $g_{r,\cO_1,\cO_2}(x)$, where $r \in \{a,f,d\}$ gives the 
direction and $x=|x_1-x_2|$ is the distance between the two points.

In the discussion of our numerical results we are a bit sloppy with the 
notation and 
use $g$ also for the numerical estimates obtained from the 
simulation. Hence there might be also a dependence on the linear 
lattice size that is not indicated explicitly.

In the case of the three-point functions 
\begin{equation}
G_{r,\cO_1,\cO_2,\cO_3}(x)=\langle \cO_1(x_1) \cO_2(x_2) 
\cO_3(x_3) \rangle
\end{equation}
we study the two choices $\cO_1 = \cO_2 = \sigma$ and $\cO_3 = \epsilon$ and 
 $\cO_1 = \cO_2 = \cO_3 = \epsilon$. Furthermore, we consider two different 
geometries that are indicated by $r$. For $r=f$ the largest displacement 
is along a face diagonal. For example
\begin{equation}
 x_3-x_1 = (j,0,0) \;\; , \; x_3-x_2 = (0,j,0) \;.
\end{equation}
Our second choice is indicated by $r=d$ and the largest displacement
is along a space diagonal.  For example
\begin{equation}
 x_3-x_1 = (j,0,0) \;\; , \; x_3-x_2 = (0,j,j) \;,
\end{equation}
where $j$ is integer and also here we sum in our simulation 
over all choices that are related by symmetry to reduce the statistical
error. The argument $x$ of $G$ gives the largest distance between two points
$x=|x_1-x_2|$. 

In order to eliminate the constants $C_i$, eq.~(\ref{twopoint}), and the power 
law behaviour from the 
three-point functions, we directly normalized our estimates of the 
three-point functions by the corresponding ones of two-point functions.
For the direction $r=f$ we get
\begin{equation}
\label{f1}
f_{\sigma \sigma \epsilon} \simeq   2^{-\Delta_{\epsilon}/2 }
 \frac{<s_{x_1} s_{x_2} s^2_{x_3} >-<s_{x_1} s_{x_2} ><s^2> }
      {<s_{x_1} s_{x_2} > [<s_{x_1}^2 s_{x_3}^2> - <s^2>^2]^{1/2}}  
\end{equation}
and 
\begin{equation}
f_{\epsilon \epsilon \epsilon} \simeq
 \frac{<s_{x_1}^2 s_{x_2}^2  s^2_{x_3}>-[<s_{x_1}^2 s_{x_2}^2 >
                                       +<s_{x_1}^2 s_{x_3}^2 >
                                       +<s_{x_2}^2 s_{x_3}^2 >]<s^2>
 +2 <s^2>^3
}
      {  [<s_{x_1}^2 s_{x_3}^2> - <s^2>^2]  \;
      [<s_{x_1}^2 s_{x_2}^2 > - <s^2>^2]^{1/2} } \;,
\end{equation}
where we used that $|x_1-x_3|=|x_1-x_2|$.  For the direction $r=d$ we
get
\begin{equation}
f_{\sigma \sigma \epsilon} \simeq 3^{-\Delta_{\epsilon}/2}
 \frac{<s_{x_1} s_{x_2} s^2_{x_3} >-<s_{x_1} s_{x_2} ><s^2> }
      {<s_{x_1} s_{x_2} > [<s_{x_2}^2 s_{x_3}^2> - <s^2>^2]^{1/2}}
\end{equation}
and 
\begin{equation}
f_{\epsilon \epsilon \epsilon} \simeq
 \frac{<s_{x_1}^2 s_{x_2}^2  s^2_{x_3}>-[<s_{x_1}^2 s_{x_2}^2 >
                                       +<s_{x_1}^2 s_{x_3}^2 >
                                       +<s_{x_2}^2 s_{x_3}^2 >]<s^2>
 +2 <s^2>^3
}
      {  [<s_{x_1}^2 s_{x_3}^2> - <s^2>^2]^{1/2}  \;
        [<s_{x_2}^2 s_{x_3}^2> - <s^2>^2]^{1/2}  \;
      [<s_{x_1}^2 s_{x_2}^2 > - <s^2>^2]^{1/2} } \;.
\end{equation}

\section{Finite size effects} 
\label{fse}
Compared with the linear size $L$ of the lattice, the distances that 
we consider for our two- and three-point functions are small. In that respect 
they can all be viewed as operators in the even channel such as the energy density.
The free energy density on a finite lattice, for a vanishing external field
behaves  as \cite{Wilson:1973jj,Fisher:1974uq,Fisher:1998kv,Pelissetto:2000ek}
\begin{equation}
 f(\beta,L) = L^{-d} \; h(L^{1/\nu} t) + f_{ns}(t) \;,
\end{equation} 
where $t=(\beta_c-\beta)/\beta_c$ is the reduced temperature, 
$h$ and $f_{ns}$ are analytic 
functions. Taking the derivative with respect to $t$ we arrive at
\begin{equation}
 E(\beta,L) =  L^{-d+1/\nu} \; \tilde h(L^{1/\nu} t) + E_{ns}(t) \;,
\end{equation}
where $\tilde h=- h'/\beta_c$. Setting $\beta=\beta_c$ we get
\begin{equation}
\label{finiteL}
 E(\beta_c,L) = L^{-d+1/\nu} \; \tilde h(0) + E_{ns}(0) \;\;, 
\end{equation}
where $d-1/\nu=d-y_t=\Delta_{\epsilon}$. 

Given the huge amount of data, we abstain from sophisticated fitting 
with Ans\"atze motivated by eq.~(\ref{finiteL}).   Instead we consider pairs
of linear lattice sizes $L_1=L$, $L_2= 2 L$ and compute 
\begin{equation}
\label{extrapol}
 G_{ex}(2 L) := G(2 L) + \frac{G(2 L) - G(L)}{2^{\Delta_{\epsilon}}-1} \;,
\end{equation}
where $G$ is the quantity under consideration.

\section{Variance reduction}
\label{variance}
\begin{figure}
\begin{center}
\includegraphics[width=14.5cm]{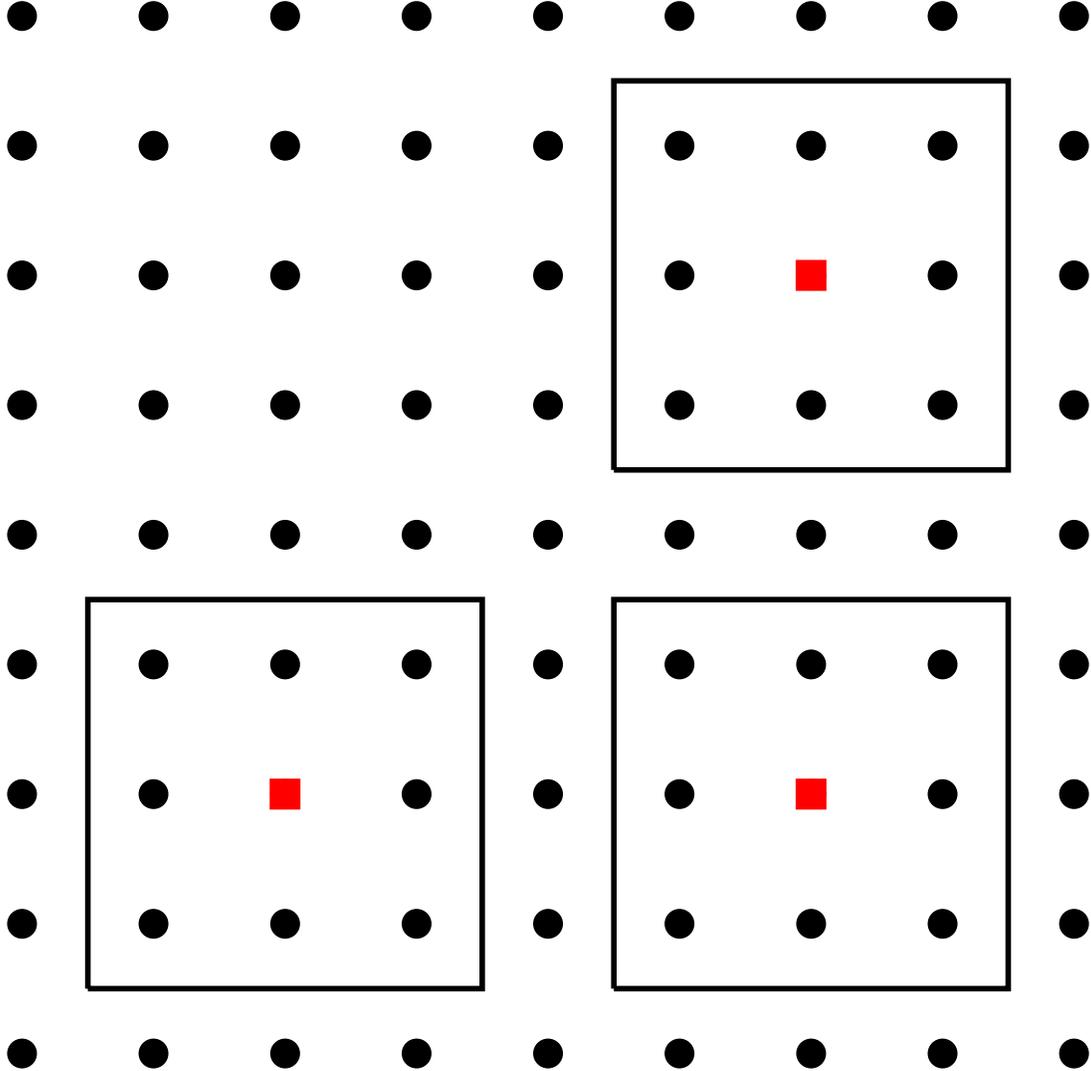}
\caption{\label{sketch} Two-dimensional sketch of the decomposition of the 
lattice 
into blocks and the remained. The three points  $x_1$, $x_2$, and $x_3$
are represented by red squares. The blocks around these three points
are bordered by solid black lines. Lattice sites different from 
$x_1$, $x_2$, or $x_3$ are given by solid black circles.
}
\end{center}
\end{figure}

Here we discuss the construction of variance reduced estimators of
$N$-point functions along the lines of refs. 
\cite{Parisi:1983hm,Luscher:2001up} in a general setting.
Let us consider the $N$-point function
\begin{equation}
 \langle \cO_1(x_1) \cO_2(x_2) ... \cO_N(x_N) \rangle
 =
\frac{\int D[\phi] \exp\left(-H[\phi]\right) \; \cO_1(x_1) \cO_2(x_2) ... \cO_N(x_N)}
     {\int D[\phi] \exp\left(-H[\phi]\right)}  \;,
\end{equation}
where $[\phi]$ denotes the collection of all fields. In the case of the 
Blume-Capel model that we simulate here, it is actually the collection of
the spins $s_x$ and  the integral becomes the sum over all configurations.
Let us consider  subsets $x_i \in \cB_i \subset \Lambda$ of sites for each $i$,
where $\Lambda$ is the set of all sites of the lattice. Typically one 
 takes $x_k \in \cB_i$  if $||x_k-x_i||\le l_{max}$ for some norm $|| \;\; ||$. 
The crucial requirement is that for any pair $i \ne j$,  no pair
of sites $x_k \in  \cB_i$ and $x_l \in  \cB_j$ exists such that 
$x_k$ and $x_l$ are nearest neighbours. In Fig. \ref{sketch} we sketch an 
implementation for a square lattice and three points. For simplicity, in the 
sketch as well as in our simulations, we use 
the Chebyshev distance. Let us denote the collection of fields living on 
$\cB_i$ by $[\phi]_i$ and the collection of fields that are not associated 
with any of the blocks by $[\phi]_R$. 
 The reduced Hamiltonian can be written 
as sum, where one summand  depends only on the fields on the remainder
and the others on the fields on the remainder and on one of the blocks:
\begin{equation}
 H[\phi] = h_R([\phi]_R) + \sum_i  h_i([\phi]_i,[\phi]_R) \;\;.
\end{equation}
This decomposition allows us to rewrite the expectation value in the 
following way:
\begin{eqnarray}
 & & \langle \cO_1(x_1) \cO_2(x_2) ... \cO_N(x_N) \rangle \nonumber \\
 &=& \frac{\int D[\phi]_R \; \exp[-h_R([\phi]_R)] \; 
\int \left(\prod_i D[\phi]_i \right) 
  \left(\prod_i  \exp[-h_i([\phi]_i,[\phi]_R)] \right) \;
   \left(\prod_i  \; \cO_i(x_i) \right)}
{\int D[\phi]_R \;  \exp[-h_R([\phi]_R)] \; \int \left(\prod_i D[\phi]_i \right) 
\left(\prod_i  \exp[-h_i([\phi]_i,[\phi]_R)] \right) } \;
      \nonumber \\
 &=& \frac{\int D[\phi]_R \; \exp[-h_R([\phi]_R)] \; \prod_i \left( \int D[\phi]_i \;
\exp[-h_i([\phi]_i,[\phi]_R)] \; \cO_i(x_i) \right)}
{\int D[\phi]_R \; \exp[-h_R([\phi]_R)] \; \prod_i \left( \int D[\phi]_i \; 
\exp[-h_i([\phi]_i,[\phi]_R)] \right) }
      \nonumber \\
 &=& \frac{\int D[\phi]_R \; \exp[-h_R([\phi]_R)] \; (\prod_i z_i([\phi]_R) ) \; 
      (\prod_i  \langle  \cO_i(x_i) \rangle_i([\phi]_R))}
{\int D[\phi]_R \;  \exp[-h_R([\phi]_R)] \; \prod_i z_i([\phi]_R) } \;,
      \nonumber \\
\end{eqnarray}
where we define
\begin{equation}
 z_i([\phi]_R) = \int D[\phi]_i \exp[-h_i([\phi]_i,[\phi]_R)] 
\end{equation}
and
\begin{equation}
\langle  \cO_i(x_i) \rangle_i([\phi]_R) 
= \frac{\int D[\phi]_i \exp[-h_i([\phi]_i,[\phi]_R)] \cO_i(x_i)}{z_i([\phi]_R)}
\;.
\end{equation}

Note that the variance of $\prod_i \langle  \cO_i(x_i) \rangle_i$ is smaller 
than that of $\prod_i \cO_i(x_i)$, since degrees of freedom have been 
integrated out. In our case we can perform an exact summation over all 
configurations if $\cB_i$ is small. In particular if $\cB_i$ contains 
only the site $x_i$ we can easily perform the summation over the 
three possible values of $s_i$. 

However it is also advantageous to perform the integration over the 
variables on the blocks $\cB_i$ approximately by using a Monte Carlo 
simulation. Let us denote the simulations restricted to field variables
on the blocks $\cB_i$ by baby simulations. These baby simulations are
started after the whole system is equilibrated. After the baby simulations 
are finished, updates on the whole system are performed. Then again
baby simulations are started. This process is iterated.
During the baby simulations, we perform $n_B$ measurements of 
$\cO_i(x_i)$ for all $i$. Let us define
\begin{equation}
 \bar{\cO_i(x_i)}_{\gamma} = \frac{1}{n_B} \sum_{\alpha=1}^{n_B}  \cO_{i,\gamma,\alpha}(x_i) \;\;,
\end{equation}
where $\alpha$ labels the measurements that are performed for a given 
configuration on the remainder. The configurations on the remainder 
are labelled by $\gamma$. We generate $n_R$ configurations on the 
remainder after 
equilibration. An estimate of the $N$-point function is obtained by
\begin{equation}
 \overline{ \overline{\cO_1(x_1)} \; \overline{\cO_2(x_2)} ...  
\overline{\cO_N(x_N)}} 
= 
\frac{1}{n_R}
\sum_{\gamma=1}^{n_R} \bar{\cO_1(x_1)}_{\gamma} \bar{\cO_2(x_2)}_{\gamma}
 ... \bar{\cO_N(x_N)}_{\gamma} \;.
\end{equation}
Let us discuss the variance of $\prod_i \bar{\cO_i(x_i)}$.
The baby Monte Carlo simulations are independent of each other. Therefore 
actually $n_B^{N}$ configurations are generated for the $N$-point 
function.  Hence, at least for small $n_b$, we expect that the 
variance of $\prod_i \bar{\cO_i(x_i)}$  behaves as $n_B^{-N}$.  As 
$n_B$ further increases the variance of
$\prod_i \langle  \cO_i(x_i) \rangle_i$ is approached. Hence the 
maximal efficiency is reached at some finite $n_B$ that in general 
has to be determined numerically.

\section{The algorithm and our implementation}
As in our previous studies of the Blume-Capel model, for example 
\cite{2016PhRvE..93c2140H}, we simulated the model 
by using a hybrid of the local heat bath algorithm,
the local Todo-Suwa algorithm \cite{ToSu13,Gutsch},   
and the single cluster algorithm \cite{Wolff:1988uh}. 
The Todo-Suwa algorithm is a local algorithm  that does not fulfil 
detailed balance but only balance. Since ergodicity can not be proved 
for the Todo-Suwa algorithm, additional local heat bath updates are 
needed to ensure ergodicity for the update scheme as a whole.
The advantage of the Todo-Suwa algorithm is that it reduces autocorrelation 
times compared with the heat bath algorithm. 

\subsection{Implementation}
We used the SIMD-oriented Fast Mersenne
Twister algorithm \cite{twister} as random number generator.  Our code
is a straight forward implementation of the algorithms in standard C.
We abstained from a parallelization of the algorithms, since in the case of the
single cluster update, a good scaling of the performance with the number
of processes is hard to achieve, see for example ref. \cite{KauRimMel10}.
We stored the value of the spins as character variable. For simplicity 
we abstain from using a more compact storage scheme, using for example two 
bits for a 
spin only. With this setup, on the hardware that is available to us,
$L=1600$ is about the largest linear lattice size that can be simulated
efficiently. In order to extrapolate in $L$ and also monitor the dependence 
of our results on the lattice size we performed simulations for the 
linear lattice sizes $L=400$, $600$, $800$, $1200$, and $1600$. 

A cycle of the update consists of one heat bath sweep followed by  $n_{cl}$
single cluster updates. Then follow $n_{ts}$ sweeps using the Todo-Suwa
algorithm.
For each of these sweeps $n_{cl}$ single cluster updates are performed.
Mostly $n_{cl}$ is chosen such that, very roughly, $n_{cl}$ times the
average cluster size equals the number of sites $L^3$.  For the range 
of lattice sizes studied here $n_{cl}=L$ is a reasonable choice.

Let us summarize an update cycle with the following pseudo-C code:
\begin{verbatim}
  heat bath sweep
  for(icl=0;icl<ncl;icl++)  single cluster update
  for(its=0;its<nts;its++)
    {
    Todo-Suwa sweep
    for(icl=0;icl<ncl;icl++)  single cluster update
    }
  baby simulations for all blocks
  measure N-point functions
\end{verbatim}

We parallelized trivially by performing several independent runs. The
equilibration takes a significant amount of CPU-time for the larger lattices.
Therefore, in the case of the linear lattice sizes $L=800$, $1200$, and $1600$,
we first equilibrated a single Markov chain. We started 
from a disordered configuration. Since in the initial phase of the simulation
clusters are small, we adapted the number of cluster updates for the first 
few update cycles, such that the aggregate cluster size per local 
update sweep is equal 
to the lattice volume or larger. After this initial phase, we continued 
with a fixed number of single cluster updates per local update sweep as 
discussed above. We performed 1000 complete update cycles to equilibrate 
the system. These update cycles are characterized by $n_{ts}=3$.
Here we only measured a few quantities like the energy density 
and the magnetisation to monitor the equilibration.
We stored the final configuration to disc. Then we started several runs
using this configuration as initial one 
and different seeds for the random number generator. In these runs we used
$n_{ts}=9$.  Since the baby simulations discussed below take a considerable
amount of CPU-time, we intend to essentially eliminate the correlation 
between subsequent measurements by using such a large number for $n_{ts}$. 
In the case of $L=800$, $1200$, and $1600$, starting from the configuration
written to disc, we performed four update cycles to
achieve a sufficient decorrelation between the branches of our simulation 
before measuring the $N$-point functions.

\subsection{The baby Monte Carlos and the  measurements} 
In our simulations we did not attempt to adjust the size of the 
blocks to the distances between the points in the $N$-point functions. 
Instead we determined the $N$-point functions for a certain range
of distances for a given decomposition into blocks.
 
In order to save CPU-time, we determined $\bar{s_x}$ and
$\bar{s_x^2}$ for a subset of the lattice sites only. 
This subset is characterized by $x^{(i)} = 0, n_s, 2 n_s, ..., L-n_s$. 
In our study we performed simulations for the strides $n_s=2$  and $4$.
We used blocks around these sites that are defined by $y \in \cB_i$ if
$|y^{(k)} - x_i^{(k)}|<n_s$ for $k=0$, $1$ and $2$. 
This means that the linear size of these blocks is
$l_b=3$ and $7$ for $n_s=2$ and $4$, respectively. Before starting the baby 
simulations we copied the spins of the block and its outer 
boundary to an auxiliary array. Since for our choice nearest neighbour blocks
overlap, we can not write all the final configurations of the baby 
simulations back to the main Markov chain. Only a subset with the larger stride 
$2 n_s$ could be used to this end. 
However, for simplicity we refrain to do so. 
At the end of the baby simulations we only stored the results 
$\bar{s_x}$ and $\bar{s_x^2}$ to compute the estimates of the $N$-point 
functions.
We computed the two- and three-point functions for the distances
$j=2 n_s$, $3 n_s$, $...$,$9 n_s$. Note that for $j=n_s$ the blocks 
overlap, and no valid result is obtained.

In the case of $n_s=2$, we performed $n_b$ sweeps over the block using the 
Todo-Suwa algorithm. Note that in the case of the baby simulations 
ergodicity is not needed. 
In preliminary simulations we varied $n_b$. It turns out that the performance
maximum is not very sharp and  depends on the type of the 
$N$-point function and on the distance $j$. Based on these experiments, 
we decided to use  $n_b=10$ in our production runs.

We performed a measurement  for the starting configuration and after each
Todo-Suwa sweep.
We reduced 
the variance by performing the sum over $s_x$ for fixed neighbours exactly:
\begin{equation}
\label{sum1}
 \tilde s_x = \frac{\sum_{s_x} \exp(\beta s_x S_x - D s_x^2) \; s_x }
                  {\sum_{s_x} \exp(\beta s_x S_x - D s_x^2)}
\end{equation}
and
\begin{equation}
\label{sum2}
 \tilde s_x^2 = \frac{ \sum_{s_x} \exp(\beta s_x S_x - D s_x^2) \; s_x^2 }
              {\sum_{s_x} \exp(\beta s_x S_x - D s_x^2)} \;,
\end{equation}
where 
\begin{equation}
S_x = \sum_{y.nn.x} s_y \;\;,
\end{equation}
where $y.nn.x$ means that $y$ is a nearest neighbour of $x$.

In the case of $n_s=4$  we used two different updating schemes. Since the 
measurement is performed  at the central site only, it might be useful to 
update the spins at central sites more often than those at the boundary.
To this end, we performed one sweep using the Todo-Suwa algorithm over
the full $7^3$ block, then follows a sweep over the central $5^3$ and 
finally a sweep over the central $3^3$ block. This sequence is repeated
$n_b$ times. A measurement using eq.~(\ref{sum1},\ref{sum2}) is performed 
for the initial configuration and after each (partial) sweep. This means
that $3 n_b+1$ measurements are performed for each baby simulation. 
After a few preliminary simulations we decided to take $n_b=30$. 
We used this scheme for our 
simulations of lattices of the linear sizes $L=400$ and $800$. 

For $L=600$, $1200$, and $1600$ we performed  cluster 
updates in addition to the sweeps with the Todo-Suwa update. 
The cluster construction is started from the fixed boundary of 
the block using the standard delete probability 
$p_d=\mbox{min}[1,\exp(-2 \beta s_x s_y)]$, where $x$ and $y$ are nearest 
neighbours. All spins that are not frozen to the boundary are flipped. 
Here we used $n_b=20$. It turned out that the cluster update gives little
advantage. However it is likely that going to larger $n_s$ this will change.
We made no attempt to construct cluster improved estimators of $s_x$ and 
$s_x^2$.

We performed simulations of the Ising model for $L=400$ and $800$ and $n_s=2$. 
In the baby Monte Carlos, we replaced the Todo-Suwa update by the Metropolis 
one. In the main Markov chain, we replaced the Todo-Suwa sweeps by 
heat bath sweeps. Note that in the case of the Ising model, the single 
cluster algorithm is ergodic and it is not necessary to add heat bath sweeps.

\subsection{Statistics of our simulations}
For a given set of parameters $n_s$ and $L$, we performed 8 up to 20 
independent runs. Each running for a month on a single core of the CPU. 
In table \ref{statis} we summarize the total number of update and measurement 
cycles  performed for each stride $n_s$ and linear lattice size $L$. 
We performed preliminary simulations without using
the variance reduction method. We abstain from discussing these
simulations in detail.

\begin{table}
\caption{\sl \label{statis}
Statistics of our simulations of the Blume-Capel model at $D=0.655$ and
$\beta=0.387721735$. For a discussion see the text.
}
\begin{center}
\begin{tabular}{rrr}
\hline
   $L$    &    $n_s=2$   &   $n_s=4$  \\
\hline
  400   &  239510 &  74270 \\
  600   &   93150 &  17810  \\
  800   &   39920 &  15670  \\
 1200   &   10300 &   1800  \\
 1600   &    4190 &   1630  \\
\hline
\end{tabular}
\end{center}
\end{table}
In  total our study took about 10 years on a single core of a recent 
CPU. 

\section{Numerical results}
Throughout we used the Jackknife method to compute statistical errors.

\subsection{The two-point functions}
\begin{figure}
\begin{center}
\includegraphics[width=14.5cm]{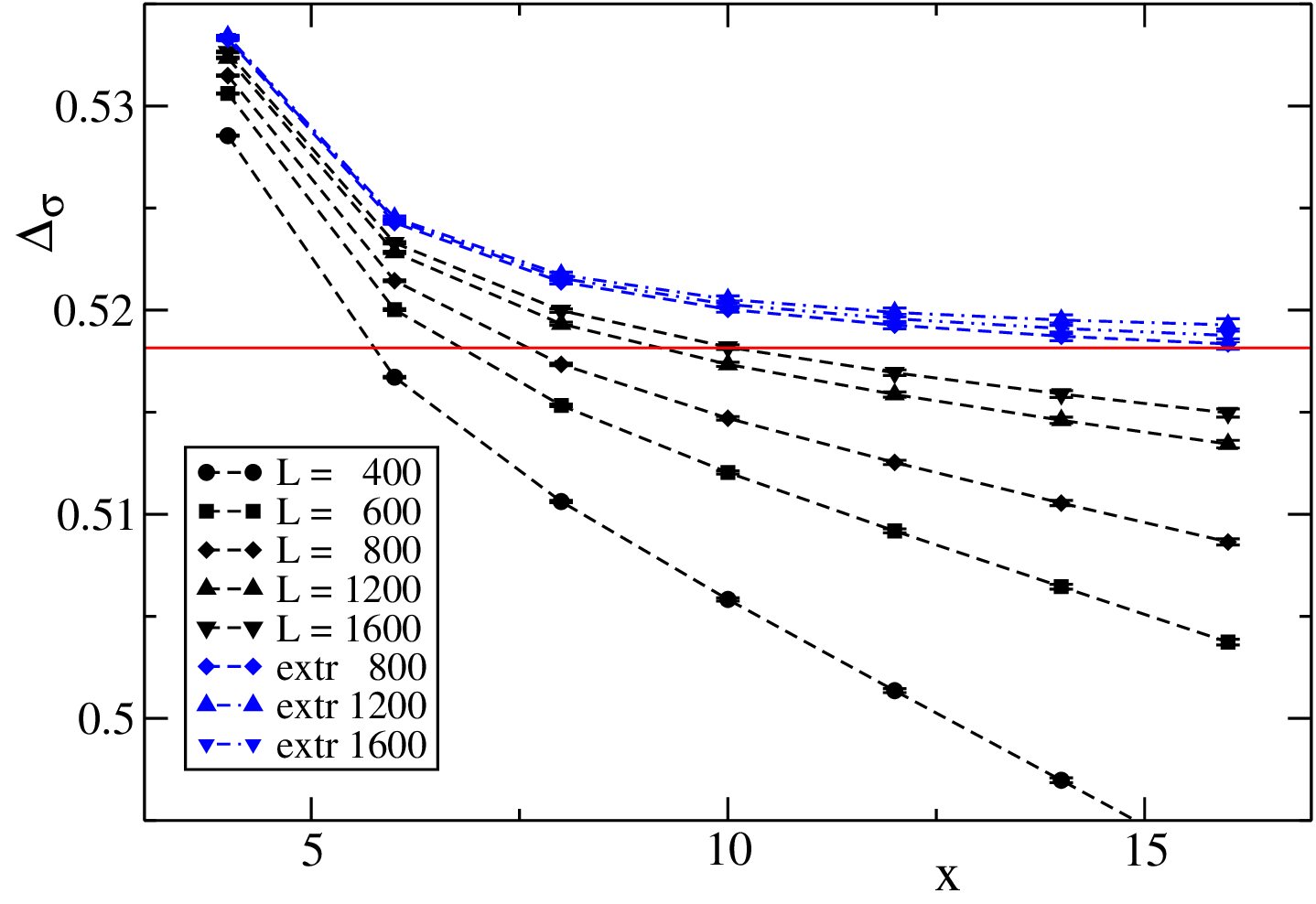}
\caption{\label{plotDs} We plot the effective dimension 
$\Delta_{\sigma,eff}(x,s)$ 
of the field as defined by eq.~(\ref{effDim}) for various linear lattice 
sizes $L$ and extrapolated estimates. The displacement between the points
is in direction $r=a$.
The dashed lines should only guide the eye.
We give only results obtained from our simulations with stride $n_s=2$ 
to keep the figure readable.  The horizontal solid line
gives $\Delta_{\sigma} =0.5181489$ obtained by the bootstrap method.
}
\end{center}
\end{figure}

First we extracted the dimension of the operators from the 
two-point functions $g(x)$. To this end, we define an effective 
exponent by
\begin{equation}
\label{effDim}
 \Delta_{eff}(x,\Delta x) = - \frac{1}{2}  \frac{\ln(g(x+\Delta x)/g(x))}
 {\ln((x+\Delta x)/x)}  \;\;,
\end{equation}
where $\Delta x=n_s$, $\Delta x= \sqrt{2} n_s$, and $\Delta x= \sqrt{3} n_s$
for $r=a$, $f$ and $d$, respectively.
Let us first discuss our results obtained for $g_{r,\sigma \sigma}$.
In Fig. \ref{plotDs} we plot results obtained for displacements along 
the axis and the stride $n_s=2$. We give results for all linear lattice sizes
that we have simulated. Up to our largest linear lattice size $L=1600$ we
see a dependence on the lattice size.  In addition we give extrapolated 
results using the pairs of lattice sizes $(400,800)$, $(600,1200)$, and 
$(800,1600)$. The extrapolated results are essentially 
consistent among 
each other, confirming the validity of the extrapolation.  It seems
plausible that the extrapolated result approaches the bootstrap value as 
$x$ increases.

\begin{figure}
\begin{center}
\includegraphics[width=14.5cm]{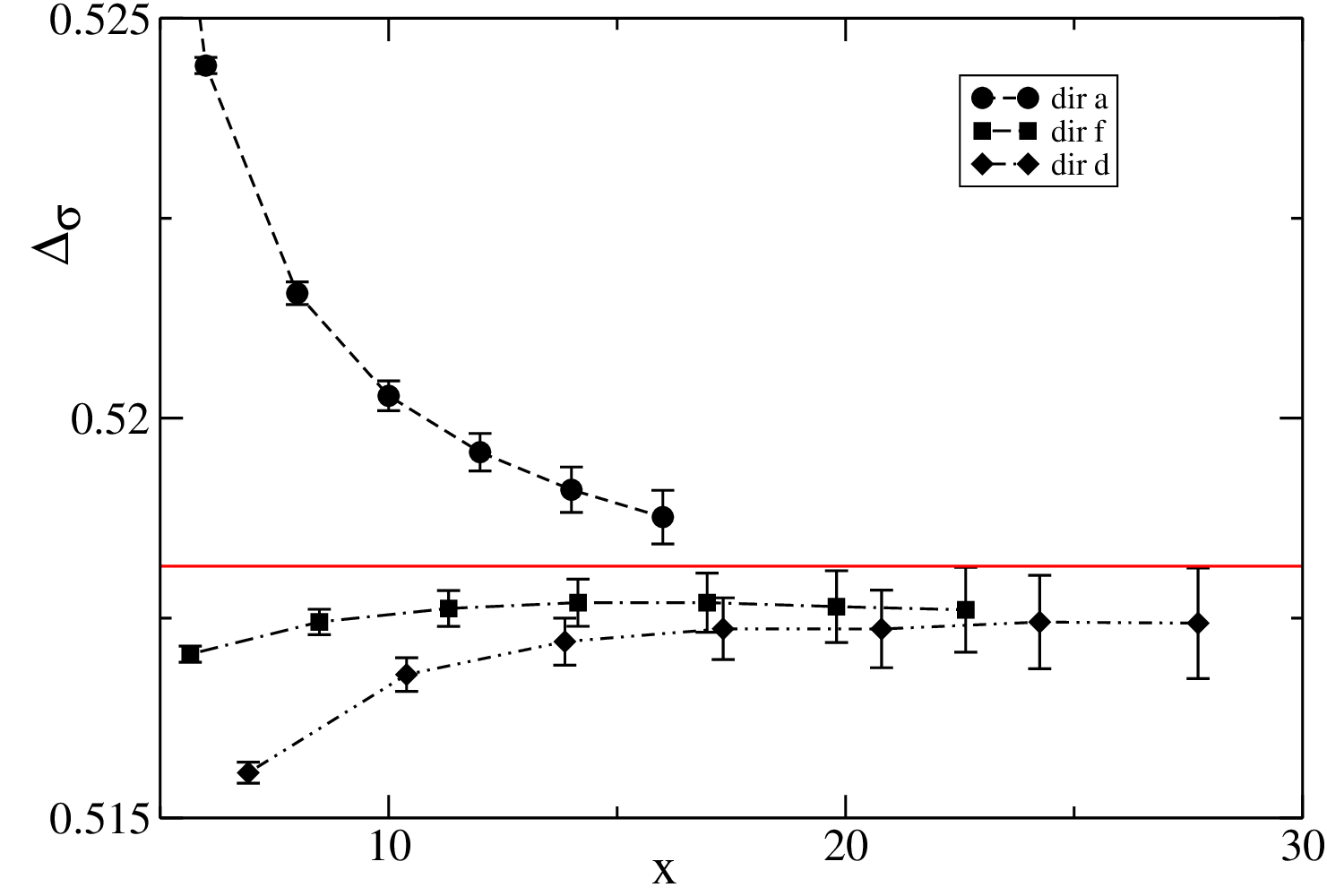}
\caption{\label{plotDsr} We plot the effective dimension
$\Delta_{\sigma,eff}(x,s)$
of the field as defined by eq.~(\ref{effDim}) for the extrapolation obtained
for $L=800$ and $1600$. All results are obtained for the stride $n_s=2$. We 
give results for all three directions of the displacement that we have studied.
The dashed lines should only guide the eye. The horizontal solid line
gives $\Delta_{\sigma} =0.5181489$ obtained by the bootstrap method.
}
\end{center}
\end{figure}

In Fig. \ref{plotDsr} we give results for the extrapolation of the pair
$(L,2L)=(800,1600)$
for all three directions that we have studied.  The small $x$ deviations
are the largest for $r=a$. For  $r=f$ and $d$ they have the 
opposite sign as for $r=a$. The amplitude of corrections is the 
smallest for $r=f$.  Since the corrections depend strongly 
on $r$, it is likely that they are dominantly caused by the
breaking of the rotational symmetry and fall off like $x^{-\omega_{NR}}$. 

Here we make no effort to extract an optimal estimate for $\Delta_{\sigma}$
from our data. 
Just looking at the figure, one might take the result for direction $f$ at 
distance $x=10 \; \sqrt{2}$ as final estimate: $\Delta_{\sigma}=0.5177(3)$. 
Using $\eta= 0.03627(10)$ \cite{Hasenbusch:2011yya}, 
we get $\Delta_{\sigma} =(1+\eta)/2=0.51814(5)$, which is consistent, but
clearly more accurate.  In the following analysis we shall use the 
bootstrap value $\Delta_{\sigma} = 0.5181489(10)$, which outpaces the 
Monte Carlo results by far. 

\begin{figure}
\begin{center}
\includegraphics[width=14.5cm]{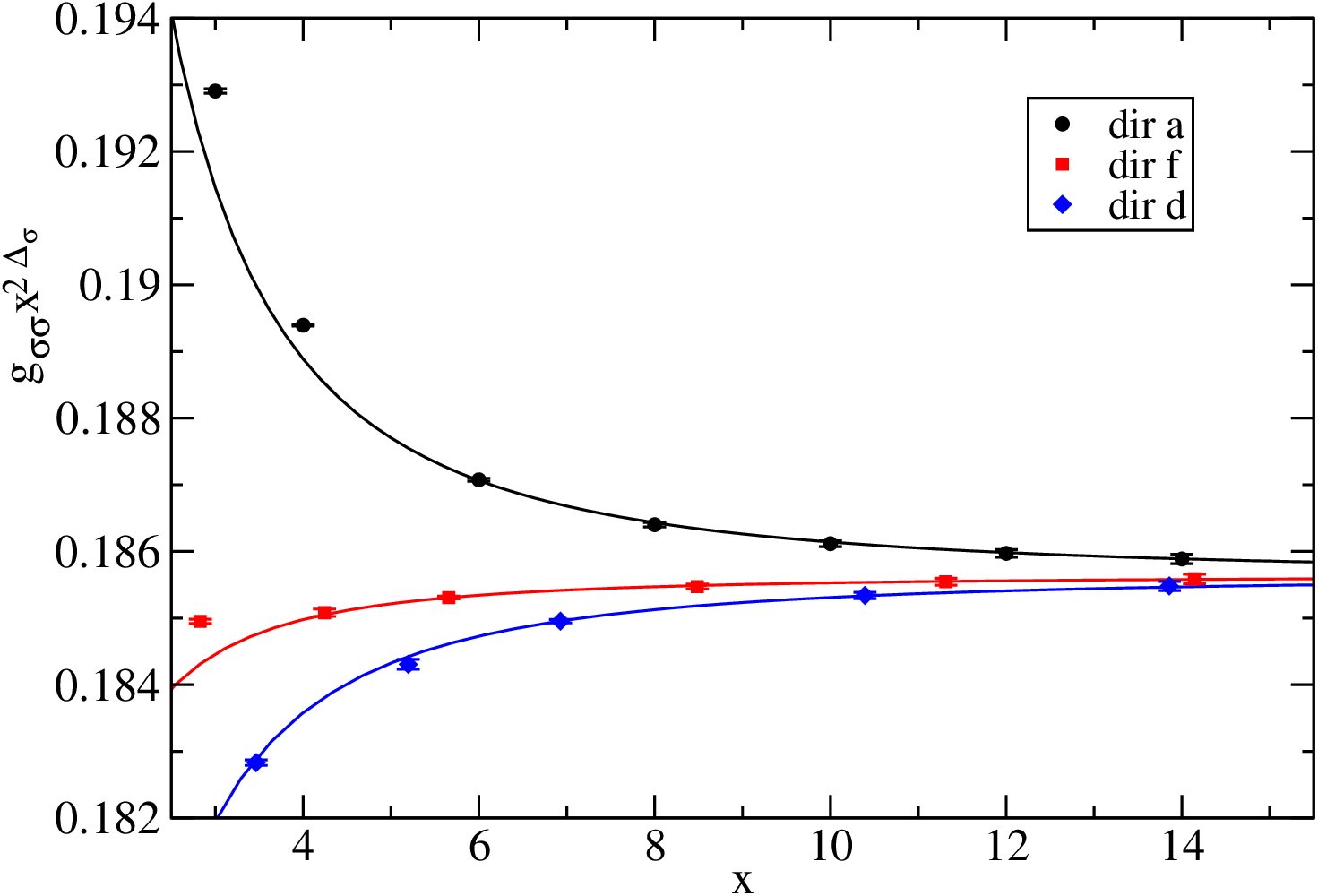}
\caption{\label{plotgss}
We plot $g_{\sigma \sigma} x^{2 \Delta_{\sigma}}$ with 
$ \Delta_{\sigma} = 0.5181489$ for the three directions studied.  
Most of the data are taken from the simulations with the stride $n_s=2$. 
A few data
at small distances are taken from our preliminary simulations without
variance reduction. The solid lines give the result of the fit with the 
Ansatz~(\ref{correctionsss}).
}
\end{center}
\end{figure}

Let us look in more detail at the corrections at small distances. Since
we study an improved model, we expect that the numerically dominant 
corrections are due to the breaking of the Galilean invariance by the 
lattice. 
In Fig. \ref{plotgss} we plot our extrapolated results for 
$g_{\sigma \sigma} x^{2 \Delta_{\sigma}}$ with 
$ \Delta_{\sigma} = 0.5181489$ for the three directions studied. 
We have fitted these data with 
\begin{equation}
\label{correctionsss}
 x^{2 \Delta_{\sigma}} g_{r,\sigma \sigma}(x) = c + a_r x^{-\omega_{NR}} 
\end{equation}
with $\omega_{NR}=2.022665$.  For simplicity, we performed a naive fit of the
 data, not
taking into account the statistical correlation of the data for
different distances and directions. We fitted jointly the data for
the three different directions, requiring that the constant $c$ is the 
same for all three directions.  The result is represented by the 
solid lines in Fig. \ref{plotgss}.  Given the shortcomings of the 
fit we do not give final results for the constants $c$ and $a_r$.
Our aim is to demonstrate that the Ansatz~(\ref{correctionsss}) indeed
describes the data well as can be seen in the plot.

\begin{figure}
\begin{center}
\includegraphics[width=14.5cm]{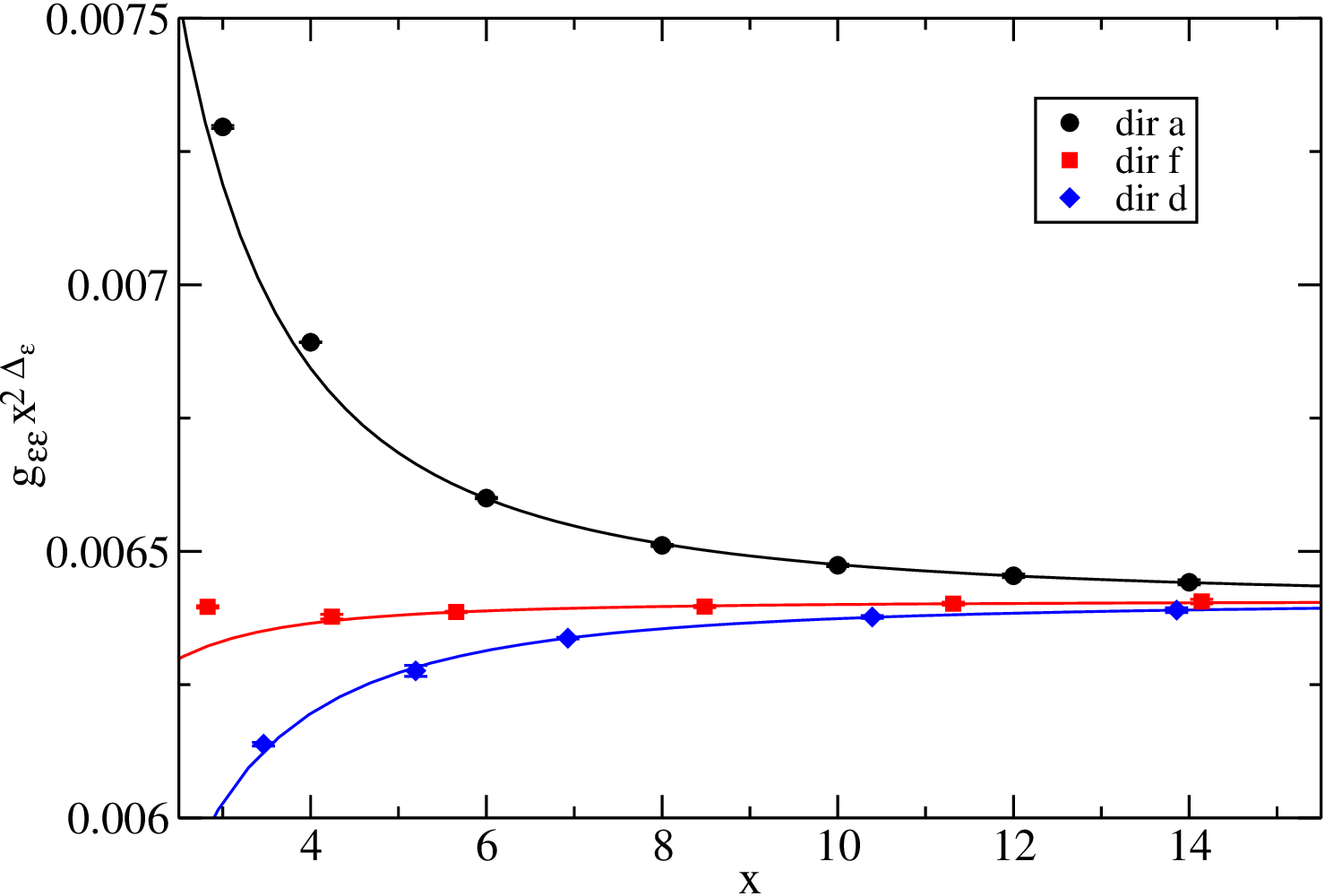}
\caption{\label{plotgee}
We plot $g_{\epsilon \epsilon} x^{2 \Delta_{\epsilon}}$ with
$ \Delta_{\epsilon} =1.412625$ for the three directions studied.
Analogous to Fig.~(\ref{plotgss}).
}
\end{center}
\end{figure}

Next we analysed the $\epsilon \epsilon$ function. 
The findings are analogous to those of the $\sigma \sigma$ 
function. Therefore we refrain from a detailed discussion.  A main difference
is that the relative statistical error increases faster as for the 
$\sigma \sigma$ function. In Fig. \ref{plotgee} we
have plotted the analogue of Fig. \ref{plotgss}. 

\subsection{The three-point functions}

\begin{figure}
\begin{center}
\includegraphics[width=14.5cm]{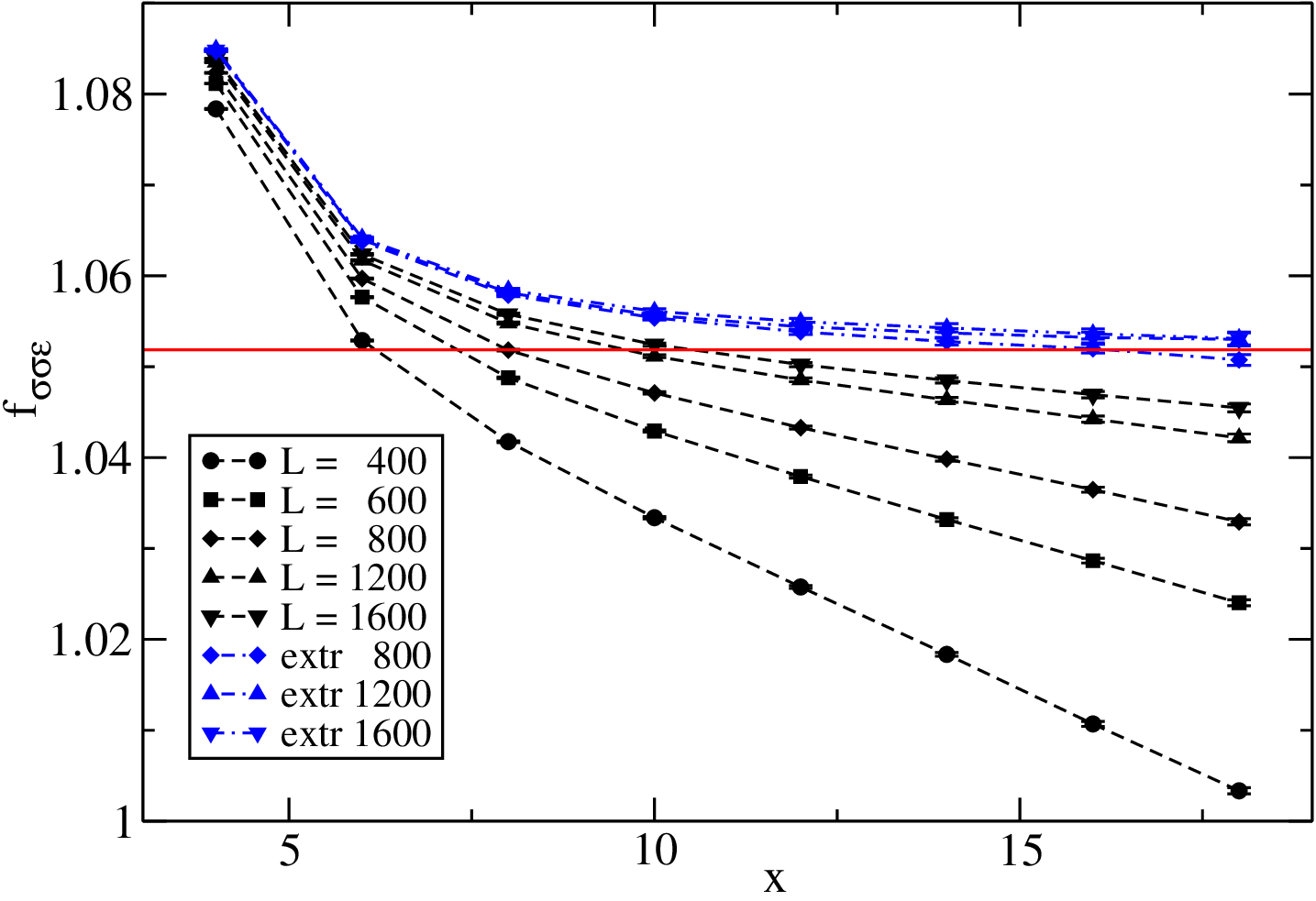}
\caption{\label{plotfsseE}
We plot our estimates of $f_{\sigma \sigma \epsilon}$ for the direction $r=f$.
We give  results obtained for the linear lattice sizes $L=400$, $600$, $800$, 
$1200$, and $1600$ and the 
extrapolations for the pairs $(L,2 L)=(400,800)$, $(600,1200)$, and 
$(800,1600)$. All data shown are based on our simulations with $n_s=2$.
The straight line gives the 
bootstrap result. 
}
\end{center}
\end{figure}

Let us first check that the extrapolation in the lattice size works. To this
end we plot in Fig. \ref{plotfsseE} our estimates of 
$f_{\sigma \sigma \epsilon}$ for $r=f$, eq.~(\ref{f1}), 
as a function of the distance $x$ for different linear lattice sizes $L$. 
In addition we give the results obtained from the 
extrapolation~(\ref{extrapol}). Note that we applied eq.~(\ref{extrapol})
to our estimates of $f_{\sigma \sigma \epsilon}$ obtained for given $L$. 
For the results obtained for a given lattice size, we see a clear dependence
on the lattice size. In contrast, the extrapolated results are consistent 
among each other.

\begin{figure}
\begin{center}
\includegraphics[width=14.5cm]{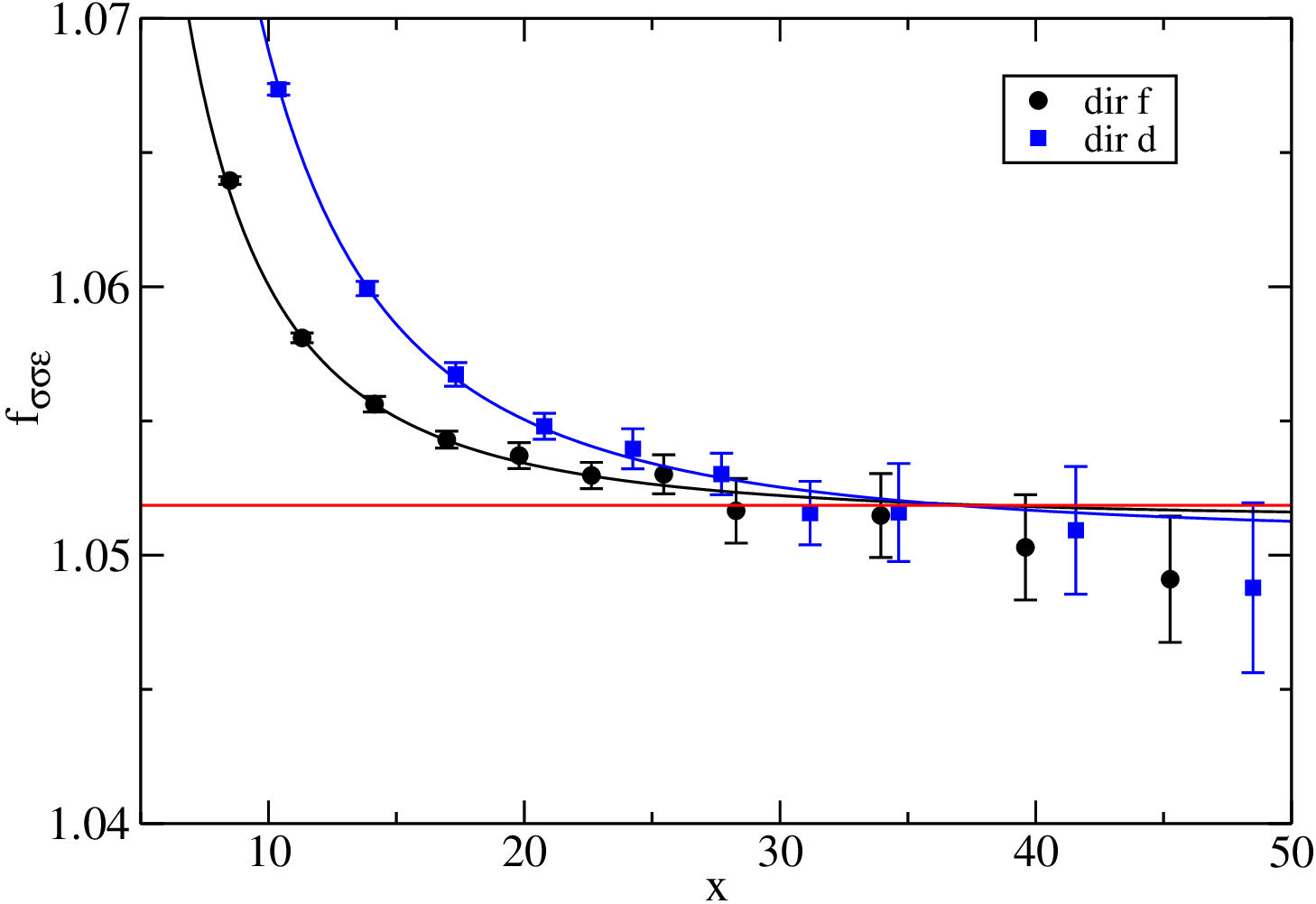}
\caption{\label{plotfess}   We plot our estimates of
$f_{\sigma \sigma \epsilon}$ as a function of the distance $x$. These
estimates are obtained by the extrapolation of the pair of lattice sizes
$(L,2L)=(800,1600)$.  The solid lines represent the results of fits with 
the Ansatz~(\ref{correctionsss}).
}
\end{center}
\end{figure}

In Fig. \ref{plotfess}  we plot our results for $f_{\sigma \sigma \epsilon}$
obtained from the extrapolation using the lattice sizes $(L,2L)=(800,1600)$. 
We give results for both geometries that we have implemented.
Here we use both the simulations with stride $n_s=2$ and $n_s=4$.  For 
distances, where data from both strides are available, we give the 
weighted average. Throughout, the estimates from $n_s=2$ are more 
accurate than those obtained from $n_s=4$.  For example for $r=f$
we get for the distance $j=16$ the estimates $1.05322(59)$ and 
$1.05242(88)$ from the simulations with stride $n_s=2$ and $n_s=4$, 
respectively.
Also here we see corrections at short distances. We fitted with 
the Ansatz~(\ref{correctionsss}). For simplicity we did not take into
account the cross correlations for different distances. We performed 
independent fits for both directions.  Good fits are obtained starting from 
$j_{min}=6$, where all  $j \ge j_{min}$ are included. 
For example, for $j_{min}=8$ we find 
$c=1.0513(3)$ and $c=1.0506(5)$ for $r=f$ and $d$, respectively. 
The errors
give only an indication, since cross correlations of the data are not taken
into account in the fit.
In Fig. \ref{plotfess} results of these fits are given as solid lines.

It is hard to quote 
a final value that is not biased by the knowledge of the bootstrap result.
But it is quite clear that our numerical data are consistent with the 
bootstrap result. Looking at Fig. \ref{plotfess} we might read off
\begin{equation}
f_{\sigma  \sigma \epsilon}= 1.051(1)
\end{equation}
from the estimate of $f_{\sigma  \sigma \epsilon}$ at distances $x\approx 30$,
without relying on the fits discussed above. 

Next we discuss our results obtained for $f_{\epsilon \epsilon \epsilon}$.
Here the statistical error increases faster with the distance than 
for  $f_{\sigma \sigma \epsilon}$. Furthermore we find that our results
obtained for the simulations with the stride $n_s=4$ are more accurate 
than those with $n_s=2$. For example for $r=f$ and $j=16$ we get
$f_{\epsilon \epsilon \epsilon} = 1.53089(85)$ and $1.53363(27)$ from 
our simulations with $n_s=2$ and $n_s=4$, respectively.
Performing fits with the Ansatz~(\ref{correctionsss}) we get 
for $j_{min}=8$ the results
$c=1.5312(14)$ and $c=1.5333(22)$ for $r=f$ and $d$, respectively.

In Fig. \ref{plotfess} the results of these fits are given as solid lines.
As final result we read off from  distances $x\approx 25$ 
\begin{equation}
 f_{\epsilon \epsilon \epsilon} = 1.533(5) 
\end{equation}
not relying on the fits with Ansatz~(\ref{correctionsss}). 

\begin{figure}
\begin{center}
\includegraphics[width=14.5cm]{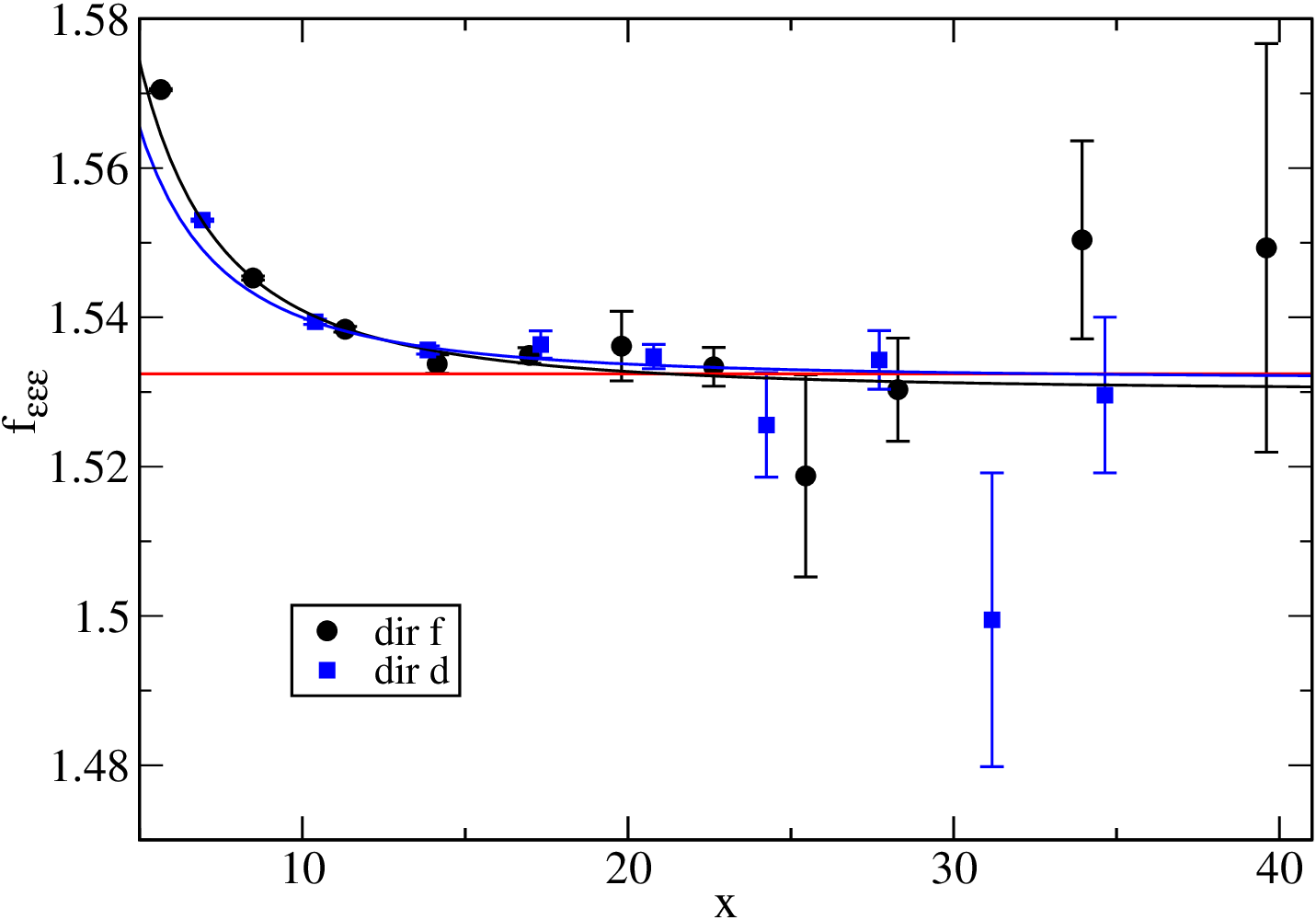}
\caption{\label{plotfeee}   We plot our estimates of 
$f_{\epsilon \epsilon \epsilon}$  as a function
of the distance $x$. The solid lines give the results of fits with the 
Ansatz~(\ref{correctionsss}).
}
\end{center}
\end{figure}

\subsection{Statistical errors}
 We find that the relative statistical error of the two-point 
 and three-point functions increases following a power law.
 The exponent is the same for $n_s=2$ and $n_s=4$ and for the 
 simulations without variance reduction. The smallest exponent 
 $\approx 1.36 $ is found for the $\sigma \sigma$ function and 
 the largest $\approx 4$ for the estimate of 
 $f_{\epsilon \epsilon \epsilon}$. This corresponds to the fact
that we see the largest gain by using the variance reduction method
in the case of $f_{\epsilon \epsilon \epsilon}$. 

\subsection{Sensitivity on the value taken as estimate of $\beta_c$}
We performed simulations for $L=800$ with stride $n_s=2$ at $\beta=0.38772$ and 
$0.38772347$ in order to estimate the dependence of the $N$-point 
functions on $\beta$. For $f_{\sigma \sigma \epsilon}$ and 
$f_{\epsilon \epsilon \epsilon}$, we find that the slope increases
with increasing distance. In the case of $f_{\sigma \sigma \epsilon}$
we find an exponent $\approx 1.25$, while for $f_{\epsilon \epsilon \epsilon}$
we get an exponent $\approx 1.38$. From the data it is not completely clear, 
whether the exponents are different.  
We estimate that the uncertainty of the estimate $\beta_c=0.387721735(25)$
used here results in  an uncertainty in the fourth digit of
$f_{\sigma \sigma \epsilon}$ and $f_{\epsilon \epsilon \epsilon}$  which
is negligible compared with the errors quoted above.

\subsection{Sensitivity on leading corrections}
Since we do not know the value of $D^*$ exactly, our results might be 
affected by residual leading corrections to scaling. 
In ref. \cite{Hasenbusch:2011yya}
we have demonstrated that the amplitude of leading corrections to scaling
should be at least reduced by a factor of 30 in the Blume-Capel model
at $D=0.655$ compared with the Ising model. Therefore we performed 
simulations of the Ising model at $\beta_c$, 
the pair of lattice sizes $L=400$ and $L=800$ and the stride $n_s=2$. 
A bit to our surprise, it is hard to see a difference between the 
results obtained from the Ising model and the improved Blume-Capel model.
We conclude that the effects of leading corrections to scaling can 
be ignored at the level of our accuracy of our final results for 
$f_{\sigma \sigma \epsilon}$ and
$f_{\epsilon \epsilon \epsilon}$.

\section{Summary and Conclusions}
We computed the operator product expansion coefficients 
$f_{\sigma \sigma \epsilon}$ and
$f_{\epsilon \epsilon \epsilon}$ by using Monte Carlo simulations
of the improved Blume-Capel model at the critical point. 
We simulated lattices with periodic
boundary conditions up to a linear size $L=1600$. We extrapolated our
results assuming that finite size corrections vanish 
$\propto L^{-\Delta_{\epsilon}}$, where $L$ is the linear size of the 
lattice.
An important ingredient of our study is the variance reduced estimator based on 
 \cite{Parisi:1983hm,Luscher:2001up}. 
In particular 
$f_{\epsilon \epsilon \epsilon}$ could be determined accurately 
due to the variance reduction.
The error of our results is smaller than that of previous estimates
obtained by Monte Carlo simulations of the three-dimensional Ising model.

Our results are fully consistent with the predictions obtained by the
conformal bootstrap method. These results are however by far more accurate 
than ours. Our study still can be understood as preliminary. Easily
more CPU time could be used and the parameters of the algorithm could 
be better tuned. However one would hardly reach the accuracy obtained 
by using the conformal bootstrap method.  The fact that our results
agree with those of the conformal bootstrap method  confirm the 
theoretical expectation that the RG fixed point that governs the 
three-dimensional Ising universality class is indeed invariant under 
conformal transformations. 

The strategy used here seems to be applicable to other universality classes.
However in the case of $O(N)$-symmetric model with $N\ge 2$ more memory 
per site is needed and therefore the linear lattice sizes that can be 
reached are smaller. It is unclear, whether these lattice sizes allow 
for a reliable extrapolation to the infinitely large system.

An interesting idea would be to combine the UV-sampler 
\cite{Herdeiro:2016agy,Herdeiro:2017jmv} 
or some similar method with the variance reduced estimator of $N$-point
functions discussed here.

\section{Acknowledgement}
This work was supported by the DFG under the grant No HA 3150/4-1.


\begin{thebibliography}{28}%
\makeatletter
\providecommand \@ifxundefined [1]{%
 \@ifx{#1\undefined}
}%
\providecommand \@ifnum [1]{%
 \ifnum #1\expandafter \@firstoftwo
 \else \expandafter \@secondoftwo
 \fi
}%
\providecommand \@ifx [1]{%
 \ifx #1\expandafter \@firstoftwo
 \else \expandafter \@secondoftwo
 \fi
}%
\providecommand \natexlab [1]{#1}%
\providecommand \enquote  [1]{``#1''}%
\providecommand \bibnamefont  [1]{#1}%
\providecommand \bibfnamefont [1]{#1}%
\providecommand \citenamefont [1]{#1}%
\providecommand \href@noop [0]{\@secondoftwo}%
\providecommand \href [0]{\begingroup \@sanitize@url \@href}%
\providecommand \@href[1]{\@@startlink{#1}\@@href}%
\providecommand \@@href[1]{\endgroup#1\@@endlink}%
\providecommand \@sanitize@url [0]{\catcode `\\12\catcode `\$12\catcode
  `\&12\catcode `\#12\catcode `\^12\catcode `\_12\catcode `\%12\relax}%
\providecommand \@@startlink[1]{}%
\providecommand \@@endlink[0]{}%
\providecommand \url  [0]{\begingroup\@sanitize@url \@url }%
\providecommand \@url [1]{\endgroup\@href {#1}{\urlprefix }}%
\providecommand \urlprefix  [0]{URL }%
\providecommand \Eprint [0]{\href }%
\providecommand \doibase [0]{http://dx.doi.org/}%
\providecommand \selectlanguage [0]{\@gobble}%
\providecommand \bibinfo  [0]{\@secondoftwo}%
\providecommand \bibfield  [0]{\@secondoftwo}%
\providecommand \translation [1]{[#1]}%
\providecommand \BibitemOpen [0]{}%
\providecommand \bibitemStop [0]{}%
\providecommand \bibitemNoStop [0]{.\EOS\space}%
\providecommand \EOS [0]{\spacefactor3000\relax}%
\providecommand \BibitemShut  [1]{\csname bibitem#1\endcsname}%
\let\auto@bib@innerbib\@empty
\bibitem [{\citenamefont {Kos}\ \emph {et~al.}(2014)\citenamefont {Kos},
  \citenamefont {Poland},\ and\ \citenamefont {Simmons-Duffin}}]{Kos:2014bka}%
  \BibitemOpen
  \bibfield  {author} {\bibinfo {author} {\bibfnamefont {F.}~\bibnamefont
  {Kos}}, \bibinfo {author} {\bibfnamefont {D.}~\bibnamefont {Poland}}, \ and\
  \bibinfo {author} {\bibfnamefont {D.}~\bibnamefont {Simmons-Duffin}},\ }\href
  {\doibase 10.1007/JHEP11(2014)109} {\bibfield  {journal} {\bibinfo  {journal}
  {JHEP}\ }\textbf {\bibinfo {volume} {11}},\ \bibinfo {pages} {109} (\bibinfo
  {year} {2014})},\ \Eprint {http://arxiv.org/abs/1406.4858} {arXiv:1406.4858
  [hep-th]} \BibitemShut {NoStop}%
\bibitem [{\citenamefont {Gliozzi}\ and\ \citenamefont
  {Rago}(2014)}]{Gliozzi:2014jsa}%
  \BibitemOpen
  \bibfield  {author} {\bibinfo {author} {\bibfnamefont {F.}~\bibnamefont
  {Gliozzi}}\ and\ \bibinfo {author} {\bibfnamefont {A.}~\bibnamefont {Rago}},\
  }\href {\doibase 10.1007/JHEP10(2014)042} {\bibfield  {journal} {\bibinfo
  {journal} {JHEP}\ }\textbf {\bibinfo {volume} {10}},\ \bibinfo {pages} {042}
  (\bibinfo {year} {2014})},\ \Eprint {http://arxiv.org/abs/1403.6003}
  {arXiv:1403.6003 [hep-th]} \BibitemShut {NoStop}%
\bibitem [{\citenamefont {Kos}\ \emph {et~al.}(2016)\citenamefont {Kos},
  \citenamefont {Poland}, \citenamefont {Simmons-Duffin},\ and\ \citenamefont
  {Vichi}}]{Kos:2016ysd}%
  \BibitemOpen
  \bibfield  {author} {\bibinfo {author} {\bibfnamefont {F.}~\bibnamefont
  {Kos}}, \bibinfo {author} {\bibfnamefont {D.}~\bibnamefont {Poland}},
  \bibinfo {author} {\bibfnamefont {D.}~\bibnamefont {Simmons-Duffin}}, \ and\
  \bibinfo {author} {\bibfnamefont {A.}~\bibnamefont {Vichi}},\ }\href
  {\doibase 10.1007/JHEP08(2016)036} {\bibfield  {journal} {\bibinfo  {journal}
  {JHEP}\ }\textbf {\bibinfo {volume} {08}},\ \bibinfo {pages} {036} (\bibinfo
  {year} {2016})},\\ \Eprint {http://arxiv.org/abs/1603.04436} {arXiv:1603.04436
  [hep-th]} \BibitemShut {NoStop}%
\bibitem [{\citenamefont
  {Simmons-Duffin}(2017{\natexlab{a}})}]{Simmons-Duffin:2016wlq}%
  \BibitemOpen
  \bibfield  {author} {\bibinfo {author} {\bibfnamefont {D.}~\bibnamefont
  {Simmons-Duffin}},\ }\href {\doibase 10.1007/JHEP03(2017)086} {\bibfield
  {journal} {\bibinfo  {journal} {JHEP}\ }\textbf {\bibinfo {volume} {03}},\
  \bibinfo {pages} {086} (\bibinfo {year} {2017}{\natexlab{a}})},\ \Eprint
  {http://arxiv.org/abs/1612.08471} {arXiv:1612.08471 [hep-th]} \BibitemShut
  {NoStop}%
\bibitem [{\citenamefont
  {Simmons-Duffin}(2017{\natexlab{b}})}]{Simmons-Duffin:2016gjk}%
  \BibitemOpen
  \bibfield  {author} {\bibinfo {author} {\bibfnamefont {D.}~\bibnamefont
  {Simmons-Duffin}},\ }in\ \href {\doibase 10.1142/9789813149441_0001} 
{\sl Proceedings, Theoretical Advanced Study Institute in Elementary}\\
    {\sl Particle
    Physics: New Frontiers in Fields and Strings (TASI 2015):
    Boulder, CO, USA, \\ June 1-26, 2015} 
(\bibinfo {year} {2017})\ pp.\
  \bibinfo {pages} {1--74},\ \Eprint {http://arxiv.org/abs/1602.07982}
  {arXiv:1602.07982 [hep-th]} \BibitemShut {NoStop}%
\bibitem [{\citenamefont {Hasenbusch}(2010)}]{Hasenbusch:2011yya}%
  \BibitemOpen
  \bibfield  {author} {\bibinfo {author} {\bibfnamefont {M.}~\bibnamefont
  {Hasenbusch}},\ }\href {\doibase 10.1103/PhysRevB.82.174433} {\bibfield
  {journal} {\bibinfo  {journal} {Phys. Rev.}\ }\textbf {\bibinfo {volume}
  {B82}},\ \bibinfo {pages} {174433} (\bibinfo {year} {2010})},\ \Eprint
  {http://arxiv.org/abs/1004.4486} {arXiv:1004.4486 [cond-mat.stat-mech]}
  \BibitemShut {NoStop}%
\bibitem [{\citenamefont {Campostrini}\ \emph {et~al.}(2002)\citenamefont
  {Campostrini}, \citenamefont {Pelissetto}, \citenamefont {Rossi},\ and\
  \citenamefont {Vicari}}]{Campostrini:2002cf}%
  \BibitemOpen
  \bibfield  {author} {\bibinfo {author} {\bibfnamefont {M.}~\bibnamefont
  {Campostrini}}, \bibinfo {author} {\bibfnamefont {A.}~\bibnamefont
  {Pelissetto}}, \bibinfo {author} {\bibfnamefont {P.}~\bibnamefont {Rossi}}, \
  and\ \bibinfo {author} {\bibfnamefont {E.}~\bibnamefont {Vicari}},\ }\href
  {\doibase 10.1103/PhysRevE.65.066127} {\bibfield  {journal} {\bibinfo
  {journal} {Phys. Rev.}\ }\textbf {\bibinfo {volume} {E65}},\ \bibinfo {pages}
  {066127} (\bibinfo {year} {2002})},\ \Eprint
  {http://arxiv.org/abs/cond-mat/0201180} {arXiv:cond-mat/0201180 [cond-mat]}
  \BibitemShut {NoStop}%
\bibitem [{\citenamefont {Butera}\ and\ \citenamefont
  {Comi}(2005)}]{Butera:2005zf}%
  \BibitemOpen
  \bibfield  {author} {\bibinfo {author} {\bibfnamefont {P.}~\bibnamefont
  {Butera}}\ and\ \bibinfo {author} {\bibfnamefont {M.}~\bibnamefont {Comi}},\
  }\href {\doibase 10.1103/PhysRevB.72.014442} {\bibfield  {journal} {\bibinfo
  {journal} {Phys. Rev.}\ }\textbf {\bibinfo {volume} {B72}},\ \bibinfo {pages}
  {014442} (\bibinfo {year} {2005})},\ \Eprint
  {http://arxiv.org/abs/hep-lat/0506001} {arXiv:hep-lat/0506001 [hep-lat]}
  \BibitemShut {NoStop}%
\bibitem [{\citenamefont {Guida}\ and\ \citenamefont
  {Zinn-Justin}(1998)}]{Guida:1998bx}%
  \BibitemOpen
  \bibfield  {author} {\bibinfo {author} {\bibfnamefont {R.}~\bibnamefont
  {Guida}}\ and\ \bibinfo {author} {\bibfnamefont {J.}~\bibnamefont
  {Zinn-Justin}},\ }\href {\doibase 10.1088/0305-4470/31/40/006} {\bibfield
  {journal} {\bibinfo  {journal} {J. Phys.}\ }\textbf {\bibinfo {volume}
  {A31}},\ \bibinfo {pages} {8103} (\bibinfo {year} {1998})}, \\  \Eprint
  {http://arxiv.org/abs/cond-mat/9803240} {arXiv:cond-mat/9803240 [cond-mat]}
  \BibitemShut {NoStop}%
\bibitem [{\citenamefont {Wilson}\ and\ \citenamefont
  {Kogut}(1974)}]{Wilson:1973jj}%
  \BibitemOpen
  \bibfield  {author} {\bibinfo {author} {\bibfnamefont {K.~G.}\ \bibnamefont
  {Wilson}}\ and\ \bibinfo {author} {\bibfnamefont {J.~B.}\ \bibnamefont
  {Kogut}},\ }\href {\doibase 10.1016/0370-1573(74)90023-4} {\bibfield
  {journal} {\bibinfo  {journal} {Phys. Rept.}\ }\textbf {\bibinfo {volume}
  {12}},\ \bibinfo {pages} {75} (\bibinfo {year} {1974})}\BibitemShut {NoStop}%
\bibitem [{\citenamefont {Fisher}(1974)}]{Fisher:1974uq}%
  \BibitemOpen
  \bibfield  {author} {\bibinfo {author} {\bibfnamefont {M.~E.}\ \bibnamefont
  {Fisher}},\ }\href {\doibase 10.1103/RevModPhys.46.597} {\bibfield  {journal}
  {\bibinfo  {journal} {Rev. Mod. Phys.}\ }\textbf {\bibinfo {volume} {46}},\
  \bibinfo {pages} {597} (\bibinfo {year} {1974})},\ \bibinfo {note} {[Erratum:
  Rev. Mod. Phys.47,543(1975)]}\BibitemShut {NoStop}%
\bibitem [{\citenamefont {Fisher}(1998)}]{Fisher:1998kv}%
  \BibitemOpen
  \bibfield  {author} {\bibinfo {author} {\bibfnamefont {M.~E.}\ \bibnamefont
  {Fisher}},\ }\bibfield  {booktitle} {\emph {\bibinfo {booktitle} {{Conceptual
  foundations of quantum field theory. Proceedings, Symposium and Workshop,
  Boston, USA, March 1-3, 1996}}},\ }\href {\doibase 10.1103/RevModPhys.70.653}
  {\bibfield  {journal} {\bibinfo  {journal} {Rev. Mod. Phys.}\ }\textbf
  {\bibinfo {volume} {70}},\ \bibinfo {pages} {653} (\bibinfo {year}
  {1998})}\BibitemShut {NoStop}%
\bibitem [{\citenamefont {Pelissetto}\ and\ \citenamefont
  {Vicari}(2002)}]{Pelissetto:2000ek}%
  \BibitemOpen
  \bibfield  {author} {\bibinfo {author} {\bibfnamefont {A.}~\bibnamefont
  {Pelissetto}}\ and\ \bibinfo {author} {\bibfnamefont {E.}~\bibnamefont
  {Vicari}},\ }\href {\doibase 10.1016/S0370-1573(02)00219-3} {\bibfield
  {journal} {\bibinfo  {journal} {Phys. Rept.}\ }\textbf {\bibinfo {volume}
  {368}},\ \bibinfo {pages} {549} (\bibinfo {year} {2002})}, \\
 \Eprint
  {http://arxiv.org/abs/cond-mat/0012164} {arXiv:cond-mat/0012164 [cond-mat]}
  \BibitemShut {NoStop}%
\bibitem [{\citenamefont {Polyakov}(1970)}]{Polyakov:1970xd}%
  \BibitemOpen
  \bibfield  {author} {\bibinfo {author} {\bibfnamefont {A.~M.}\ \bibnamefont
  {Polyakov}},\ }\href@noop {} {\bibfield  {journal} {\bibinfo  {journal} {JETP
  Lett.}\ }\textbf {\bibinfo {volume} {12}},\ \bibinfo {pages} {381} (\bibinfo
  {year} {1970})},\ \bibinfo {note} {[Pisma Zh. Eksp. Teor.
  Fiz.12,538(1970)]}\BibitemShut {NoStop}%
\bibitem [{\citenamefont {Caselle}\ \emph {et~al.}(2015)\citenamefont
  {Caselle}, \citenamefont {Costagliola},\ and\ \citenamefont
  {Magnoli}}]{Caselle:2015csa}%
  \BibitemOpen
  \bibfield  {author} {\bibinfo {author} {\bibfnamefont {M.}~\bibnamefont
  {Caselle}}, \bibinfo {author} {\bibfnamefont {G.}~\bibnamefont
  {Costagliola}}, \ and\ \bibinfo {author} {\bibfnamefont {N.}~\bibnamefont
  {Magnoli}},\ }\href {\doibase 10.1103/PhysRevD.91.061901} {\bibfield
  {journal} {\bibinfo  {journal} {Phys. Rev.}\ }\textbf {\bibinfo {volume}
  {D91}},\ \bibinfo {pages} {061901} (\bibinfo {year} {2015})},\\
 \Eprint
  {http://arxiv.org/abs/1501.04065} {arXiv:1501.04065 [hep-th]} \BibitemShut
  {NoStop}%
\bibitem [{\citenamefont {Costagliola}(2016)}]{Costagliola:2015ier}%
  \BibitemOpen
  \bibfield  {author} {\bibinfo {author} {\bibfnamefont {G.}~\bibnamefont
  {Costagliola}},\ }\href {\doibase 10.1103/PhysRevD.93.066008} {\bibfield
  {journal} {\bibinfo  {journal} {Phys. Rev.}\ }\textbf {\bibinfo {volume}
  {D93}},\ \bibinfo {pages} {066008} (\bibinfo {year} {2016})},\ \Eprint
  {http://arxiv.org/abs/1511.02921} {arXiv:1511.02921 [hep-th]} \BibitemShut
  {NoStop}%
\bibitem [{\citenamefont {Herdeiro}(2017)}]{Herdeiro:2017jmv}%
  \BibitemOpen
  \bibfield  {author} {\bibinfo {author} {\bibfnamefont {V.}~\bibnamefont
  {Herdeiro}},\ }\href {\doibase 10.1103/PhysRevE.96.033301} {\bibfield
  {journal} {\bibinfo  {journal} {Phys. Rev. E}\ }\textbf {\bibinfo {volume}
  {96}},\ \bibinfo {pages} {033301} (\bibinfo {year} {2017})},\ \Eprint
  {http://arxiv.org/abs/1705.11045} {arXiv:1705.11045 [cond-mat.stat-mech]}
  \BibitemShut {NoStop}%
\bibitem [{\citenamefont {Herdeiro}\ and\ \citenamefont
  {Doyon}(2016)}]{Herdeiro:2016agy}%
  \BibitemOpen
  \bibfield  {author} {\bibinfo {author} {\bibfnamefont {V.}~\bibnamefont
  {Herdeiro}}\ and\ \bibinfo {author} {\bibfnamefont {B.}~\bibnamefont
  {Doyon}},\ }\href {\doibase 10.1103/PhysRevE.94.043322} {\bibfield  {journal}
  {\bibinfo  {journal} {Phys. Rev.}\ }\textbf {\bibinfo {volume} {E94}},\
  \bibinfo {pages} {043322} (\bibinfo {year} {2016})},\\ \Eprint
  {http://arxiv.org/abs/1605.05350} {arXiv:1605.05350 [cond-mat.stat-mech]}
  \BibitemShut {NoStop}%
\bibitem [{\citenamefont {Parisi}\ \emph {et~al.}(1983)\citenamefont {Parisi},
  \citenamefont {Petronzio},\ and\ \citenamefont {Rapuano}}]{Parisi:1983hm}%
  \BibitemOpen
  \bibfield  {author} {\bibinfo {author} {\bibfnamefont {G.}~\bibnamefont
  {Parisi}}, \bibinfo {author} {\bibfnamefont {R.}~\bibnamefont {Petronzio}}, \
  and\ \bibinfo {author} {\bibfnamefont {F.}~\bibnamefont {Rapuano}},\ }\href
  {\doibase 10.1016/0370-2693(83)90930-9} {\bibfield  {journal} {\bibinfo
  {journal} {Phys. Lett.}\ }\textbf {\bibinfo {volume} {128B}},\ \bibinfo
  {pages} {418} (\bibinfo {year} {1983})}\BibitemShut {NoStop}%
\bibitem [{\citenamefont {L{\"u}scher}\ and\ \citenamefont
  {Weisz}(2001)}]{Luscher:2001up}%
  \BibitemOpen
  \bibfield  {author} {\bibinfo {author} {\bibfnamefont {M.}~\bibnamefont
  {L{\"u}scher}}\ and\ \bibinfo {author} {\bibfnamefont {P.}~\bibnamefont
  {Weisz}},\ }\href {\doibase 10.1088/1126-6708/2001/09/010} {\bibfield
  {journal} {\bibinfo  {journal} {JHEP}\ }\textbf {\bibinfo {volume} {09}},\
  \bibinfo {pages} {010} (\bibinfo {year} {2001})},\ \Eprint
  {http://arxiv.org/abs/hep-lat/0108014} {arXiv:hep-lat/0108014 [hep-lat]}
  \BibitemShut {NoStop}%
\bibitem [{\citenamefont {Deng}\ and\ \citenamefont
  {Bl{\"o}te}(2004)}]{DeBl04}%
  \BibitemOpen
  \bibfield  {author} {\bibinfo {author} {\bibfnamefont {Y.}~\bibnamefont
  {Deng}}\ and\ \bibinfo {author} {\bibfnamefont {H.~W.~J.}\ \bibnamefont
  {Bl{\"o}te}},\ }\href {\doibase 10.1103/PhysRevE.70.046111} {\bibfield
  {journal} {\bibinfo  {journal} {Phys. Rev. E}\ }\textbf {\bibinfo {volume}
  {70}},\ \bibinfo {pages} {046111} (\bibinfo {year} {2004})}\BibitemShut
  {NoStop}%
\bibitem [{\citenamefont {Hasenbusch}(2001)}]{myhabil}%
  \BibitemOpen
  \bibfield  {author} {\bibinfo {author} {\bibfnamefont {M.}~\bibnamefont
  {Hasenbusch}},\ }\href {\doibase 0.1142/S0129183101002383} {\bibfield
  {journal} {\bibinfo  {journal} {Int. J. Mod. Phys. C}\ }\textbf {\bibinfo
  {volume} {12}},\ \bibinfo {pages} {911} (\bibinfo {year} {2001})}\BibitemShut
  {NoStop}%
\bibitem [{\citenamefont {Hasenbusch}(2012)}]{Hasenbusch:2012spc}%
  \BibitemOpen
  \bibfield  {author} {\bibinfo {author} {\bibfnamefont {M.}~\bibnamefont
  {Hasenbusch}},\ }\href {\doibase 10.1103/PhysRevB.85.174421} {\bibfield
  {journal} {\bibinfo  {journal} {Phys. Rev. B}\ }\textbf {\bibinfo {volume}
  {85}},\ \bibinfo {pages} {174421} (\bibinfo {year} {2012})},\ \Eprint
  {http://arxiv.org/abs/1202.6206} {arXiv:1202.6206 [cond-mat.stat-mech]}
  \BibitemShut {NoStop}%
\bibitem [{\citenamefont {{Hasenbusch}}(2016)}]{2016PhRvE..93c2140H}%
  \BibitemOpen
  \bibfield  {author} {\bibinfo {author} {\bibfnamefont {M.}~\bibnamefont
  {{Hasenbusch}}},\ }\href {\doibase 10.1103/PhysRevE.93.032140} {\bibfield
  {journal} {\bibinfo  {journal} {\pre}\ }\textbf {\bibinfo {volume} {93}},\
  \bibinfo {eid} {032140} (\bibinfo {year} {2016})},\ \Eprint
  {http://arxiv.org/abs/1512.02491} {arXiv:1512.02491 [cond-mat.stat-mech]}
  \BibitemShut {NoStop}%
\bibitem [{\citenamefont {Todo}\ and\ \citenamefont {Suwa}(2013)}]{ToSu13}%
  \BibitemOpen
  \bibfield  {author} {\bibinfo {author} {\bibfnamefont {S.}~\bibnamefont
  {Todo}}\ and\ \bibinfo {author} {\bibfnamefont {H.}~\bibnamefont {Suwa}},\
  }\href {\doibase 10.1088/1742-6596/473/1/012013} {\bibfield  {journal}
  {\bibinfo  {journal} {J. Phys.: Conf. Ser.}\ }\textbf {\bibinfo {volume}
  {473}},\ \bibinfo {pages} {012013} (\bibinfo {year} {2013})},\\ \Eprint
  {http://arxiv.org/abs/1310.6615} {arXiv:1310.6615 [cond-mat.stat-mech]}
  \BibitemShut {NoStop}%
\bibitem{Gutsch}
F. Gutsch, {\sl Markov-Ketten ohne detailliertes Gleichgewicht}, Bachelor thesis, Humboldt-Universit\"at
zu Berlin (2014).
\bibitem [{\citenamefont {Wolff}(1989)}]{Wolff:1988uh}%
  \BibitemOpen
  \bibfield  {author} {\bibinfo {author} {\bibfnamefont {U.}~\bibnamefont
  {Wolff}},\ }\href {\doibase 10.1103/PhysRevLett.62.361} {\bibfield  {journal}
  {\bibinfo  {journal} {Phys. Rev. Lett.}\ }\textbf {\bibinfo {volume} {62}},\
  \bibinfo {pages} {361} (\bibinfo {year} {1989})}\BibitemShut {NoStop}%
\bibitem{twister}
M. Saito and M. Matsumoto,
``SIMD-oriented Fast Mersenne Twister:
a 128-bit Pseudorandom Number Generator'',
in
{\sl Monte Carlo and Quasi-Monte Carlo Methods 2006},
edited by A. Keller, S. Heinrich, H. Niederreiter, (Springer, 2008);
M. Saito, Masters thesis, Math. Dept., Graduate School of science,
Hiroshima University, 2007.
The source code of the program is provided at
``http://www.math.sci.hiroshima-u.ac.jp/$\sim$m-mat/MT/SFMT/index.html''
\bibitem [{\citenamefont {Kaupu$\breve{\mbox{z}}$s}\ \emph
  {et~al.}(2010)\citenamefont {Kaupu$\breve{\mbox{z}}$s}, \citenamefont
  {Rim$\breve{\mbox{s}}$$\bar{\mbox{a}}$ns},\ and\ \citenamefont
  {Melnik}}]{KauRimMel10}%
  \BibitemOpen
  \bibfield  {author} {\bibinfo {author} {\bibfnamefont {J.}~\bibnamefont
  {Kaupu$\breve{\mbox{z}}$s}}, \bibinfo {author} {\bibfnamefont
  {J.}~\bibnamefont {Rim$\breve{\mbox{s}}$$\bar{\mbox{a}}$ns}}, \ and\ \bibinfo
  {author} {\bibfnamefont {R.~V.~N.}\ \bibnamefont {Melnik}},\ }\href {\doibase
  10.1103/PhysRevE.81.026701} {\bibfield  {journal} {\bibinfo  {journal} {Phys.
  Rev. E}\ }\textbf {\bibinfo {volume} {81}},\ \bibinfo {pages} {026701}
  (\bibinfo {year} {2010})}\BibitemShut {NoStop}%
\end{thebibliography}
\end{document}